\documentclass[preprintnumbers,amsmath,11pt,amssymb,floatfix,superscriptaddress,nofootinbib]{revtex4}
\usepackage{graphicx}
\usepackage{epsfig}
\usepackage{bm}
\usepackage{amsfonts}

\begin{document}

\title{\textbf{Pilgrim Dark Energy in $f(T, T_G)$ cosmology}}

\author{Surajit Chattopadhyay}
\footnote{Corresponding author}
\email{surajitchatto@outlook.com}
\affiliation{Pailan College of Management and Technology, Bengal
Pailan Park, Kolkata-700 104, India.}

\author{Abdul Jawad}
\email{jawadab181@yahoo.com}
\affiliation{Department of Mathematics, Lahore Leads University,
Lahore, Pakistan.}

\author{Davood Momeni}
\email{d.momeni@yahoo.com}
\affiliation{Eurasian International Center for Theoretical Physics and Department of General
Theoretical Physics, Eurasian National University, Astana 010008, Kazakhstan.}

\author{Ratbay Myrzakulov}
\email{rmyrzakulov@gmail.com}
\affiliation{Eurasian International Center for Theoretical Physics and Department of General
Theoretical Physics, Eurasian National University, Astana 010008, Kazakhstan.}

\date{\today}

\begin{abstract}
We work on the reconstruction scenario of pilgrim
dark energy" (PDE) in $f(T, T_G)$. In PDE
model it is assumed that a repulsive force that is accelerating the Universe is phantom type
with $(w_{DE}<-1)$ and it so strong that prevents  formation of the black hole. We construct the
$f(T, T_G)$ models and correspondingly evaluate equation of state
parameter for various choices of scale factor. Also, we assume
polynomial form of $f(T, T_G)$ in terms of cosmic time and
reconstruct $H$ and $w_{DE}$ in this manner. Through discussion, it
is concluded that PDE shows aggressive phantom-like behavior for
$s=-2$ in $f(T, T_G)$ gravity.
\end{abstract}

\maketitle

\section{Introduction}
Accelerated expansion of the current universe is well established
through observational studies (Perlmutter et al. 1999; Bennett et
al. 2003; Spergel et al. 2003; Tegmark et al. 2004; Abazajian et al.
2004, 2005; Allen et al. 2004). It is believed that this expansion
is due to missing energy component, also dubbed as Dark Energy (DE)
characterized by negative pressure. Reviews on DE include
Padmanabhan (2005), Copeland et al. (2006), Li et al. (2011) and
Bamba et al. (2012a), Nojiri and Odintsov (2007). The $\Lambda$CDM model, the simplest DE
candidate, is consistent very well with all observational data.
However, it has the following two weak points as enlisted in
Nesseris et al. (2013): The $\Lambda$ has fine tuning amount and good  marginal
adoption with  cosmological observations in large scales.
This motivates the researchers in proposing the wide range of more
complex generalized cosmological models of DE which have been
discussed in Copeland et al.(2006) and Bamba et al. (2012a,2012b).

In recent years, the holographic dark energy (HDE) (Li, 2004,2011), is
based on holographic Universe idea and it is one of the interesting and powerful candidates for the DE.  and its density is given by (Li, 2004)
\begin{eqnarray}
\rho_{\Lambda}=3c^2m_p^2L^{-2}
\end{eqnarray}
where $L$ is the IR cutoff, $m_p=(8\pi G)^{-1/2}$ is the reduced
Planck mass.
 IR cut-offs $L$ parameter is
consideredin various ways in numerious articles: s the Hubble horizon $H^{-1}$, particle horizon,
the future event horizon, the Ricci scalar curvature radius. Also it was proposed a generalization of holographic models in which the cut-off parameter has a more general form  (S. ’i. Nojiri and S. D. Odintsov, 2006a). The last generalized holographic dark energy scenario has been investigated from different point of the views,specially the stability under small perturbations. In all of these models,  the cut-off length scale
is  proportional to the causal length that considered to the
perturbations of the flat spacetime .
Further more, it is suggested that phenomenon of matter collapse can be avoided
through the existence of appropriate repulsive force. In the present
set up of cosmological scenario, this can be only prevented through
phantom-like DE which contains enough repulsive force. Different
attempts have been made in this way, e.g., reduction of BH mass
though phantom accretion phenomenon (Babichev et al. (2004);
Martin-Moruno 2008; Jamil et al. 2008; Babichev et al. 2008; Jamil
2009; Jamil and Qadir 2011; Sharif and Abbas 2011;Jamil et al. 2011; Bhadra and
Debnath 2012; Sharif and Jawad 2013) and the avoidance of event
horizon in the presence of phantom-like DE (Lobo 2005a, 2005b;
Sushkov 2005, Sharif and Jawad 2014).

Moreover, it is predicted that phantom DE with strong negative
pressure can push the universe towards the big rip singularity where
all the physical objects lose the gravitational bounds and finally
dispersed. This prediction supports the phenomenon of the avoidance
of BH formation and motivated Wei (2012) in constructing PDE model.
He pointed out different possible theoretical and observational ways
to make the BH free phantom universe with Hubble horizon through PDE
parameter. Further, Sharif and Jawad (2013a, 2013b, 2014) have
analyzed this proposal in detail by chosing different IR cutoffs
through well-known cosmological parameters in flat and non-flat
universes. This model has also been in different modified gravities
(Sharif and Rani (2014); Sharif and Zubair (2014)).
 Another direction that one can follow to
explain the acceleration is to modify the gravitational sector
itself, acquiring a modified cosmological dynamics. However, note
that apart from the interpretation, one can transform from one
approach to the other, since the crucial issue is just the number of
degrees of freedom beyond General Relativity and standard model
particles (Kofinas et al. 2014; Sahni and Starobinsky, 2006).
Detailed review in this modified gravity approach is available in
references like Nojiri and Odintsov (2007), Tsujikawa (2010) ,Cai et al.(2009) and
Clifton et al. (2012). In the majority of modified gravitational
theories, one suitably extends the curvature based Einstein-Hilbert
action of General Relativity.
However, an interesting class of gravitational modification arises
when one modifies the action of the equivalent formulation of
General Relativity based on torsion (Kofinas and Saridakis, 2014).
Inspired by the $f(R)$ modifications of the Einstein-Hilbert
Lagrangian, $f(T)$ modified gravity has been proposed by extending
$T$ to an arbitrary function (Ferraro and Fiorini,2007). Diffrent aspects of the  $f(T)$ has been investigated in the literature    (Aslam  et al. 2013; Bamba et al. (2012a,2012b,2013a,2013b,2013c,2014a,2014b,2014c,2014d); Farooq et al. 2013; Houndjo et al. (2012, 2013); Jamil et al. (2013a,2013b,2012a-2012i); Cai et al. 2011; Momeni et al.(2011,2012);Rodrigues et al. (2013a,2013b); Setare  et al.  2011) .
The simplest modified gravity is obtained by replacing $R\to f(R)$,is so called $f(R)$ gravity. This kind of the modified gravities have several interesting extensions(for a recent review see (Nojiri and Odintsov (2014)). The next modification is inspired from the string theory. It comes from the Gauss- Bonnet term $G$ and widely investigated in the literature (Capozziello et al. 2014; Nojiri and Odintsov 2011a,2011b,2011c; Nojiri et al. 2010; Bamba et al. 2010a,2010b; Cognola et al. 2009; Capozziello et al. 2009;Capozziello et al. 2013;  Bamba et al. 2008; Nojiri et al. 2007b;  Cognola et al. 2007,2008; Nojiri and Odintsov 2007a,c,d,2006e; Nojiri at al. 2006b,2006c,; Cognola et al. 2006a,2006b; Nojiri and Odintsov  2005a,2005b,2005c,2006d; Nojiri et al. 2005a,2005b; Nojiri et al. 2002; Lidsey et al. 2002; Cvetic at al. 2002; Noji et al. 2009;Nojiri and Odintsov 2008; Nojiri and Odintsov 2003 ).

The main motivation to consider GB models is they are motivated from the string theory. In low limit of string theory these higher order curvature corrections appered. Motivated by $f(R)$ gravity,we can introduce the $f(G)$ gravity ,proposed by Nojiri and Odintsov (2005). This modification of the Einstein gravity unified dark matter and dark energy (Cognola et al. 2006) in a same scenario and in a consistent way and also in relation to the gauge/gravity proposal(Lidsey et al. 2002a,2002b).
To include the higher GB terms in f(T) gravity and motivated from the $f(R,G)$ model,
recently  $f(T, T_G)$ has been
constructed on the basis of $T$ (old quadratic torsion scalar) and
$T_G$ (new quartic torsion scalar $T_G$ that is the teleparallel
equivalent of the Gauss-Bonnet term) (Kofinas and Saridakis, 2014).
This theory also belongs to a novel class of gravitational
modification. Cosmological applications for this gravity have also
been presented in detailed.
To realize the role of DE in modified gravity,a very useful technique proposed by (Nojiri and Odintsov 2006e,2005a, 2006d) and it extended for several cosmological scenarios. Consider the first Friedmann equation of a type of modified gravity in the following form:
\begin{eqnarray}
\kappa^2\rho_{DE}=\Sigma_{\Sigma n_{i}=n} A(R,G,...)\frac{\partial^n f(R,G,...)}{\partial R^{n_1} \partial G^{n_2}...}
\end{eqnarray}
In the above equation , $f(R,G,...)$ stands for the modified gravity action. For a model of DE,for example the holographic DE, in general $\rho_{DE}=\rho_{DE}(H,\dot{H},..)$. Also, implicitly we able to write it as $\rho_{DE}=\rho_{DE}(R,G,...)$. So, if we can solve the following partial differential equations for $f(R,G,...)$, indeed we reconstructed the modified gravity for this type of DE. Also, the reconstruction scheme works if we assume that in any cosmological epoch ,$a(t)$ is given (for a review see Nojiri and Odintsov 2011c). Our aim in this work is to reconstruct $f(T,T_G)$ for PDE in flat FRW Universe.

Here, we also check cosmological aspects of this theory in flat FRW
universe. It is interesting to mention that we provide the
correspondence scenario of newly proposed gravity theory as well as
dynamical DE model (PDE). \par
The plan of the paper is given as follows: Next
section contains the reconstruction scheme. We provide the
construction of reconstructed $f(T, T_G)=\tilde{f}$ models and
corresponding EoS parameter with respect to PDE parameter $(s)$. In
section \textbf{(II)}, we have reported a reconstruction approach
through power-law form of scale factor. In section \textbf{(III)},
we present reconstruction scheme with a choice of Hubble rate $H$
leading to unification of matter and dark energy dominated universe.
In section
\textbf{(IV)}, we choose the scale factor in the ``intermediate"
form and reconstruct $f$ and subsequently $w_{DE}$. A bouncing scale
factor in power law form is considered in section  \textbf{(V)} and
a reconstruction scheme is reported. An analytic form of $f$ is
assumed in section \textbf{(VI)} and Hubble parameter is
reconstructed without any choice of scale factor.  The paper is
concluded in section \textbf{(VII)}.

\section{Reconstruction scheme for $f(T,T_G)$ through power-law scale factor}
A new and valid generalization of $f(T)$ in the modified gravity models  as $f(T,T_G)$  based on T and equivalent to Gauss-Bonnet term $T_G$ in the
 teleparallel , is quite different from their counterparts  $f(T)$ and $f(R,G)$ in Einstein gravity (Kofinas et
al., 2014). In $f(T,T_G)$ gravity
\begin{eqnarray}
S =\frac{1}{2}\!\int d^{4}\!x\,e\,F(T,T_G)+\!\int d^{4}\!x\,e\,\mathcal{L}_m\,,
\end{eqnarray}
where in “God-given
natural units” $c=1,\ \ \kappa^2\equiv 8\pi G=1$  and $\mathcal{L}_m$ is Lagrangian of the matter fields. For  a spatially flat
cosmological  FRW metric
\begin{equation}
ds^{2}=-
dt^{2}+a^{2}(t)\Big(d\vec{x}. d\vec{x}\Big).
\end{equation}
We obtain:
\begin{eqnarray}\label{TTG}
T=6H^2~~~~~~~~~~~~~~~~\\
T_G=24H^2(\dot{H}+H^2)
\end{eqnarray}
where $H=\dot{a}/a$ is the Hubble parameter and dots denote
differentiation with respect to $t$. Friedmann equations in the
usual form are
\begin{eqnarray}
H^2=\frac{1}{3}(\rho_m+\rho_{DE})~~~~~~~~~~~~~~~\label{friedman1}\\
\dot{H}=-\frac{1}{2}(\rho_m+\rho_{DE}+p_m+p_{DE})\label{friedman2}
\end{eqnarray}
Kofinas et al. (2014) modified the Eqs. (\ref{friedman1}) and
(\ref{friedman2}) by defining the energy density and pressure of the
effective dark energy sector as
\begin{eqnarray}
\rho_{DE}=\frac{1}{2}\left(6H^2-f+12H^2f_T+T_G~f_{T_G}-24H^3\dot{f}_{T_G}\right)\label{modifiedrho}\\
p_{DE}=\frac{1}{2}\left[-2 (2\dot{H}+3H^2)+f-4(\dot{H}+3H^2)f_T-4H\dot{f}_T-Tf_{T_G}+\right.\nonumber\\
\left.\frac{2}{3H}T_G\dot{f}_{T_G}+8H^2\ddot{f}_{T_G}\right]\label{modifiedp}
\end{eqnarray}
Standard matter and dark energy are conserved separately, i.e. the evolution equations are
\begin{eqnarray}
\dot{\rho}_m+3H(\rho_m+p_m)=0\label{mattercons}\\
\dot{\rho}_{DE}+3H(\rho_{DE}+p_{DE})=0\label{energycons}
\end{eqnarray}
 The first property of PDE is
\begin{eqnarray}\label{pde1}
\rho_{\Lambda}\gtrsim m_p^2 L^{-2}
\end{eqnarray}
To implement Eq. (\ref{pde1}), Wei (2012) set PDE in the simplest way as
\begin{eqnarray}\label{pde}
\rho_{\Lambda}=3n^2 m_p^{4-s}L^{-s}
\end{eqnarray}
where $n$ and $s$ are both dimensionless constants. From Eqs.
(\ref{pde1}) and (\ref{pde}) we have $L^{2-s}\gtrsim
m_p^{s-2}=l_p^{2-s}$, where $l_p$ is the reduced Plank length. Since
$L>l_p$, one requires
\begin{eqnarray}\label{sless2}
s \leq 2
\end{eqnarray}
Another requirement for PDE is that it is to be phantom-like (Wei, 2012) i.e.
\begin{eqnarray}\label{wpde}
w_{\Lambda}<-1
\end{eqnarray}
It was stated in Wei (2012) that to obtain the EoS for PDE, we should choose a particular cut-off $L$ and the simplest cut-off  is the Hubble horizon $L=H^{-1}$.

The PDE mentioned above would now be studied in $f(T,T_G)$ gravity
proposed recently by Kofinas et al. (2014).
As we are going to consider PDE in $f(T,T_G)$ gravity in Eq. (\ref{modifiedrho}) we use ($L=1/H$)
\begin{eqnarray}\label{eql}
\rho_{DE}=\rho_{\Lambda}=3n^2 m_p^{4-s}\left(\frac{t}{m}\right)^{-s}
\end{eqnarray}
Reconstruction scheme is a way to solve the following second order partial differential equation for $f(T,T_G)$:
\begin{eqnarray}
&&\rho_{DE}(T,T_G) =\frac{T}{2}-\frac{1}{2}f(T,T_G)
+T{\frac {\partial }{\partial T}}f(T,T_G) +\frac{1}{2}\,G{\frac {
\partial }{\partial G}}f(T,T_G)\\&&\nonumber -\frac{1}{2}\,A \left( T,T_G
 \right) {\frac {\partial ^{2}}{\partial {G}^{2}}}f \left( T,T_G
 \right) -\frac{1}{2}\,B \left( T,T_G \right) {\frac {\partial ^{2}}{\partial T
\partial T_G}}f \left( T,T_G \right)
\end{eqnarray}
There is no simple solution for this partial differential equation for a given set of functions $\{A,B\}$. But sometimes for a choice of scale factor $a(t)$ we can solve it.\par
We consider the scale factor in the form
\begin{equation}\label{a}
a(t)=a_0 t^m
\end{equation}
where $m>0$. Subsequently Hubble parameter $H$ and its time derivative $\dot{H}$ are
\begin{eqnarray}\label{H}
H=\frac{m}{t},~~~\dot{H}=-\frac{m}{t^2}
\end{eqnarray}
\begin{eqnarray}\label{T}
T=\frac{6 m^2}{t^2},~~
T_G=\frac{24 (-1+m) m^3}{t^4};~~
\dot{T}=-\frac{12 m^2}{t^3};~~
\dot{T}_G=-\frac{96 (-1+m) m^3}{t^5}
\end{eqnarray}
Considering Eqs. (\ref{modifiedrho}), (\ref{eql}) and (\ref{T}) we get the following differential equation
\begin{eqnarray}\label{diffeqn}
\left(\frac{t^{2+s} }{4-4
m}\right)\frac{d^2f}{dt^2}+t^{1+s}\frac{5}{4}\left(\frac{m-2}{m-1}\right)
\frac{df}{dt}+t^s f-6 m^2 t^{-2+s}+6 m^s n^2=0
\end{eqnarray}
This differential equation has an exact solution given by the following expression:
\begin{eqnarray}
&&f (t) ={C_2}{t}^{\frac{5}{2}m-\frac{9}{2}+\frac{1}{2}\sqrt {65-74m+25{m}^{2}}}
+{C_1
}{t}^{\frac{5}{2}m-\frac{9}{2}-\frac{1}{2}\sqrt {65-74m+25{m}^{2}}}+\\&& \nonumber{\frac {120\left( -\frac{8}{5}\,{m}^{1+s}+\frac{3}{5}\,{m}^{2+s}+{m}^{s}
 \right) {n}^{2}{t}^{2-s}-60\, \left(  \left( s-\frac{4}{5} \right) m +\frac{1}{5}\,{s}
^{2}+\frac{4}{5}-\frac{9}{5}\,s \right)  \left( m-1 \right) {m}^{2}}{15{t}^{2} \left(
 \left( s-\frac{4}{5} \right) m+\frac{1}{5}\,{s}^{2}+\frac{4}{5}-\frac{9}{5}\,s \right)  \left( -\frac{5}{3}+m
 \right) }}
\end{eqnarray}
It is possible to write $f(T,T_G)$ in the following form:
\begin{eqnarray}
&&f(T,T_G) = C_2\left( {\frac {2{T}^{3/2}\sqrt {6}}{3\,G-2\,{T}^{2}}} \right) ^{
{\frac {-5{T}^{2}}{3\,T_G-2\,{T}^{2}}}-9/2+1/2\sqrt {65+{\frac {148{
T}^{2}}{3T_G-2{T}^{2}}}+{\frac {100{T}^{4}}{ \left( 3T_G-2\,{T}^{2
} \right) ^{2}}}}}\\&&\nonumber+ C_1\left( {\frac {2{T}^{3/2}\sqrt {6}}{3
T_G-2\,{T}^{2}}} \right) ^{{\frac {-5{T}^{2}}{3T_G-2\,{T}^{2}}}-9/2-
1/2\sqrt {65+{\frac {148{T}^{2}}{3T_G-2{T}^{2}}}+{\frac {100{T
}^{4}}{ ( 3T_G-2{T}^{2}) ^{2}}}}} \\&&\nonumber-32805
( T_G-2/3{T}^{2}) ^{4} (( T_G-2/3\,{T}^{2}
) ^{4}[ 3/5 ( \,{\frac {-2{T}^{2}}{3T_G-2\,{T}^{2}
}} ) ^{2+s}\\&&\nonumber-8/5\left( {\frac {-2{T}^{2}}{3T_G-2\,{T}^{2}}}
\right) ^{1+s}+ \left( {\frac {-2{T}^{2}}{3T_G-2{T}^{2}}}
\right) ^{s} ] {n}^{2}\left( 2{\frac {{T}^{3/2}\sqrt {6}}{3
T_G-2\,{T}^{2}}} \right) ^{2-s}\\&&\nonumber+{\frac {2}{45}}(( -2/3
{s}^{2}+8/3s ) {T}^{2}+T_G( -9s+4+{s}^{2}
) {T}^{4}T_G ) ( 3T_G-2{T}^{2}) ^{-4}\\&&\nonumber\times
( 8{T}^{2}s+3{s}^{2}T_G-2{s}^{2}{T}^{2}+12T_G-27sT_G
) ^{-1}( -4{T}^{2}+15T_G ) ^{-1}{T}^{-3}
\end{eqnarray}

Based on the choice of the scale factor we have $\dot{H}<0$ is valid
in the whole cosmic history. Hence, considering
$w_{\Lambda}=-1-\frac{s\dot{H}}{3H^2}<-1$ as required by PDE, we
need $s<0$. It was clearly established in Wei (2012) that for PDE
\begin{itemize}
  \item EoS $w_{\Lambda}$ goes asymptotically to $ -1$ in the late;
  \item $w_{\Lambda}<-1$;
  \item $w_{\Lambda}$ never crosses the phantom divide $w = −1$ in the whole cosmic history.
\end{itemize}
In order to verify whether consideration of PDE in the $f(T,T_G)$
gravity, which is a class of modified gravity, leads to results
consistent with that of Wei (2012) we consider both $0<s\leq~2$ as
well as $s<0$. In particular we consider $s=2$ and $s=-2$. Solving
Eq. (\ref{diffeqn}) for the said two cases we have two solutions for
$f(t)$ as:
\begin{eqnarray}\nonumber
\tilde{f}(s=2)&=&\frac{12(-1+m)m^2\left(-1+n^2\right)}{(-5+3m)
t^2}+t^{\frac{1}{2}\left(-9+5m-\sqrt{65+m(-74+25
m)}\right)}\left[~C_2+C_1\right.\\\label{f2}&\times&\left.t^{\sqrt{65+m
(-74+25m)}}\right]
\end{eqnarray}
and
\begin{eqnarray}\nonumber
\tilde{f}(s=-2)&=&\frac{12(-1+m)\left(\frac{m^4}{5-3m}+\frac{n^2
t^4}{13-7m}\right)}{m^2t^2}+t^{\frac{1}{2}\left(-9+5
m+\sqrt{65+m(-74+25m)}\right)}\left[C_3+C_4
\right.\\\label{f-2}&\times&\left.t^{-\sqrt{65+m(-74+25m)}}\right]
\end{eqnarray}
It may be noted that from now onwards $\tilde{f}$ would denote the reconstructed $f$. Using solution (\ref{f2}) in Eq. (\ref{modifiedp}) we get the modified $p_{DE}$ for $s=2$ as function of $t$
\begin{eqnarray}\nonumber
p_{DE}(s=2)&=&\frac{1}{192(-1+m)m(-5+3m)}t^{\frac{1}{2}(-11-\xi)}
\left[m^5\left(-1728n^2t^{\frac{7+\xi}{2}}-750C_1t^{\frac{5
m}{2}+\xi}\right)\right.\\\nonumber &+&\left.384 m^4
t^{\frac{5+\xi}{2}}\left(-2+4t+n^2(2+11t)\right)-192m^3t^{\frac{5+\xi}{2}}\left(-8+16t+n^2(8+15t)\right)\right.\\\nonumber
&+&\left.25C_1m^4t^{\frac{5 m}{2}+\xi}(158+45t-18\xi)+
+5C_1t^{\frac{5
m}{2}+\xi}(-9+\xi)\left(-6-53t+4(2\right.\right.\\\nonumber
&+&\left.\left.t)\xi+(-2+t)\xi^2\right)+3C_1m^3t^{\frac{5m}{2}+\xi}(t(-1884+65
\xi)-10(254+3(-22\right.\\\nonumber&+&\left.\xi)\xi))-C_1
mt^{\frac{5m}{2}+\xi
}(2382-2\xi(962+\xi(-179+8\xi))+t(8396+\xi(-957\right.\\\nonumber
&+&\left.\xi(-50+3\xi))))-C_2(-5+3m)t^{5m/2}\left(250m^4-75m^3(5t+2(6+\xi))\right.\right.\\\nonumber
&-&\left.\left.(9+\xi)(-2(1+\xi)(3+\xi)+t(-53+(-4+\xi
)\xi))+m^2(t(1259+65\xi)\right.\right.\\\nonumber
&+&\left.\left.10(104+\xi(41+3\xi)))+m(-2(6+\xi)(37+\xi(22+\xi))+t(-1393\right.\right.\\\nonumber
&+&\left.\left.\xi(-138+7\xi)))\right)
+m^2\left(384t^{\frac{5+\xi}{2}}\left(-2+4t+n^2(2+t)\right)-C_1t^{\frac{5
m}{2}+\xi}\right.\right.\\\label{p-1}
&\times&\left.\left.(-6532+2\xi(1532+3(-53+\xi)\xi)+t(-10474\xi(739+21\xi)))\right)\right]
\end{eqnarray}

Using solution (\ref{f-2}) in Eq. (\ref{modifiedp}) we get the
modified $p_{DE}$ for $s=-2$ as function of $t$
\begin{eqnarray}\nonumber
p_{DE}(s=-2)&=&\frac{t^{\frac{1}{2}(-11-\xi)}}{192(-1+m)m^3(-5+3m)
(-13+7m)}\left[-5250C_3m^8t^{\frac{5 m}{2}+\xi}+25344
m^5t^{\frac{5+\xi}{2}}\right.\\\nonumber&\times&
\left.(-1+2t)-20736m^6t^{\frac{5+\xi}{2}}(-1+2t)+5376m^7t^{\frac{5+\xi}{2}}(-1+2t)+5760n^2t^{\frac{13+\xi}{2}}\right.\\\nonumber
&\times&\left.(-10+7t)-576mn^2t^{\frac{13+\xi}{2}}(-260+97t)
-192m^4t^{\frac{5+\xi}{2}}\left(-52+104t+63n^2t^5\right)\right.\\\nonumber&+&
\left.25 C_3 m^7 t^{\frac{5
m}{2}+\xi}(1496+315t-126\xi)+m^2\left(-576n^2t^{\frac{13+\xi}{2}}(220+17t)-65C_3\right.\right.\\\nonumber&\times&
\left.\left.t^{\frac{5m}{2}+\xi}(-9+\xi)\left(-6-53t+4(2+t)\xi+(-2+t)\xi^2\right)\right)+C_3m^6t^{\frac{5m}{2}+\xi}(3t\right.\\\nonumber&\times&
\left.(-18063+455\xi)-10(10469+9\xi(-219+7\xi)))-C_3m^5t^{\frac{5m}{2}+\xi}(-144784\right.\\\nonumber&+&
\left.2\xi(23594+3\xi(-566+7\xi))+t(-146794+\xi(7708+147\xi)))-C_3m^4t^{\frac{5
m}{2}+\xi}\right.\\\nonumber&\times&
\left.(10(10159+\xi(-5330+(664-19\xi)\xi))+t(194934+\xi(-16306+7\xi(-89\right.\\\nonumber&+&
\left.3\xi))))-C_4
m^2(65+m(-74+21m))t^{5m/2}\left(250m^4-75m^3(5t+2(6+\xi))\right.\right.\\\nonumber&-&
\left.\left.(9+\xi)(-2(1+\xi)(3+\xi)+t(-53+(-4+\xi)\xi))+m^2(t(1259+65\xi)+10\right.\right.\\\nonumber&\times&
\left.\left.(104+\xi(41+3\xi)))+m(-2(6+\xi)(37+\xi(22+\xi))+t(-1393+\xi(-138\right.\right.\\\nonumber&+&
\left.\left.+7\xi)))\right)+m^3\left(2880n^2t^{\frac{13+\xi}{2}}(12+13t)+C_3t^{\frac{5
m}{2}+\xi}(32856-2\xi(13871+\xi\right.\right.\\\label{p-2}&\times&
\left.\left.(-2782+139
\xi))+t(125843+\xi(-15556+\xi(-825+74\xi))))\right)\right]
\end{eqnarray}

where $\xi=\sqrt{65+m (-74+25 m)}$. Using Eqs. (\ref{p-1}) and
(\ref{p-2}) the EoS for PDE i.e. $w_{DE}=\frac{p_{DE}}{3n^2
m_p^{4-s}\left(\frac{t}{m}\right)^{-s}}$ is obtained through
reconstructed $\tilde{f}$ for $s=2$ and $s=-2$ respectively as

\begin{eqnarray}\nonumber
w_{DE}(s=2)&=&\frac{1}{72m^3(-5+3m)n^2}t^{-\frac{1}{2}
\left(7+\xi\right)}\left[130C_1\xi t^{\frac{5m}{2}+\xi}-333C_1m\xi
t^{\frac{5m}{2}+\xi}+278C_1m^2\xi\right.\\\nonumber&\times& \left.
t^{\frac{5m}{2}+\xi}-75C_1m^3\xi t^{\frac{5m}{2}+\xi}+15C_1\xi
t^{1+\frac{5m}{2}+\xi}-34C_1m\xi t^{1+\frac{5m}{2}+\xi}+15C_1m^2\xi
\right.\\\nonumber&=&\left.t^{1+\frac{5m}{2}+\xi}+C_2(-5+3m)t^{5
m/2}\left(218-125m^3+26\xi+\left(19+3\xi\right)t+5m^2\right.\right.\\\nonumber&-&
\left.\left.(88+5\xi+5t)-m\left(533+51\xi+\left(52+5\xi\right)
t\right)\right)+C_1(-5+3m)t^{\frac{5m}{2}+\xi}
(218\right.\\\nonumber&+& \left.19t+m(-13(41+4t)+5m(88-25m+5t)))-24
m^2t^{\frac{1}{2}\left(5+\xi\right)}(-4+8t\right.\\&+&\left.9m^2n^2
t+2n^2(2+t)-m\left(-4+8t+n^2(4+13t)\right))\right] \label{w2}
\end{eqnarray}
and
\begin{eqnarray}\nonumber
w_{DE}(s=-2)&=&\frac{1}{72 m \xi^2 n^2}t^{\frac{1}{2}
\left(-15-\xi\right)}\left[-1690 C_3 m^2 \xi t^{\frac{5
m}{2}+\xi}+5239 C_3 m^3 \xi t^{\frac{5 m}{2}+\xi}-5945 C_3 m^4
\right.\\\nonumber&\times& \left.\xi t^{\frac{5 m}{2}+\xi}+2921 C_3
m^5 \xi t^{\frac{5 m}{2}+\xi}-525 C_3 m^6 \xi t^{\frac{5
m}{2}+\xi}-195 C_3 m^2 \xi t^{1+\frac{5 m}{2}+\xi}+547
\right.\\\nonumber&\times& \left.C_3 m^3 \xi t^{1+\frac{5
m}{2}+\xi}-433 C_3 m^4 \xi t^{1+\frac{5 m}{2}+\xi}+105 C_3 m^5 \xi
t^{1+\frac{5 m}{2}+\xi}-24 t^{\frac{1}{2} \left(5+\xi\right)}
\right.\\\nonumber&\times& \left.\left(4 (-1+m) m^4 (-13+7 m)-8
(-1+m) m^4 (-13+7 m) t-60 (5+m (-8\right.\right.\\\nonumber&+&
\left.\left.3 m))n^2t^4+3 (-5+3 m) (-14+m (-3+7 m)) n^2
t^5\right)-C_3 m^2 (65+m(-74\right.\\\nonumber&+& \left.21 m))
t^{\frac{5 m}{2}+\xi}(-218-19 t+m (533+5 m (-88+25 m-5 t)+52 t))-C_4
m^2 \right.\\\nonumber&\times&\left.(65+m(-74+21m))t^{5m/2}
\left(-218+125m^3-26\xi-\left(19+3\xi\right)
t-5m^2(88\right.\right.\\\label{w-2}&+&\left.\left.5\xi+5
t)+m\left(533+51\xi+\left(52+5\xi\right)t\right)\right)\right]
\end{eqnarray}
Firstly we include the discussion of reconstructed $f$ models
corresponding to PDE parameter $s=2,~-2$, respectively. We plot
$f(s=2,-2)$ versus cosmic time as well as cosmic scale factor $m>1$
as shown in FIG. \textbf{1} and \textbf{2}. In FIG.\textbf{1}, it is
observed that $\tilde{f}(s=2)$ shows decreasing behavior from very high
value and approaches to zero versus cosmic time in the range $2\leq
m\leq2.3$. However, for $m>2.3$, it shows decreasing behavior
initially, becomes flat for a short interval of time, and then
exhibits increasing behavior. FIG. \textbf{2} indicates that
$\tilde{f}(s=-2)$ increases with cosmic time from very low value, becomes
flat for a glimpse of time interval and then decreases for $2\leq
m\leq2.2$. For $2.2<m\leq2.5$, it increases but approaches to zero
after short interval of time. For $2.5<m$, $f(s=-2)$ increases with
cosmic time from very low value, becomes flat for a glimpse of time
interval and then increases.
\begin{figure}[ht] \begin{minipage}[b]{0.45\linewidth}
\centering\includegraphics[width=\textwidth]{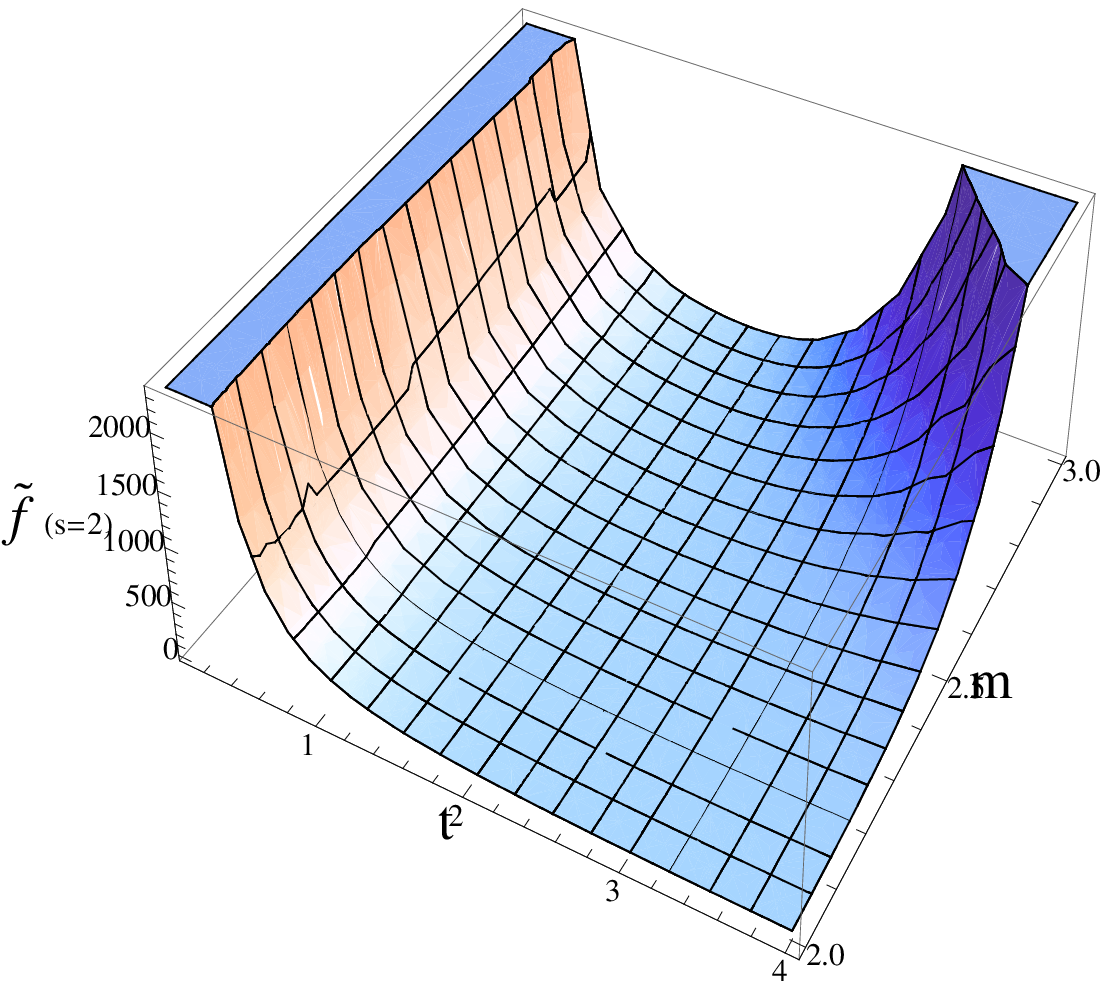}
\caption{Plot of reconstructed $f$ (Eq. (\ref{f2})) from PDE taking
$s=2$ and $m>1$. Also, $n=3,~~C_1=0.5,~~C_2=0.2$.}
\label{M1restrict} \end{minipage} \hspace{0.5cm} \hspace{0.5cm}
\begin{minipage}[b]{0.45\linewidth}
\centering\includegraphics[width=\textwidth]{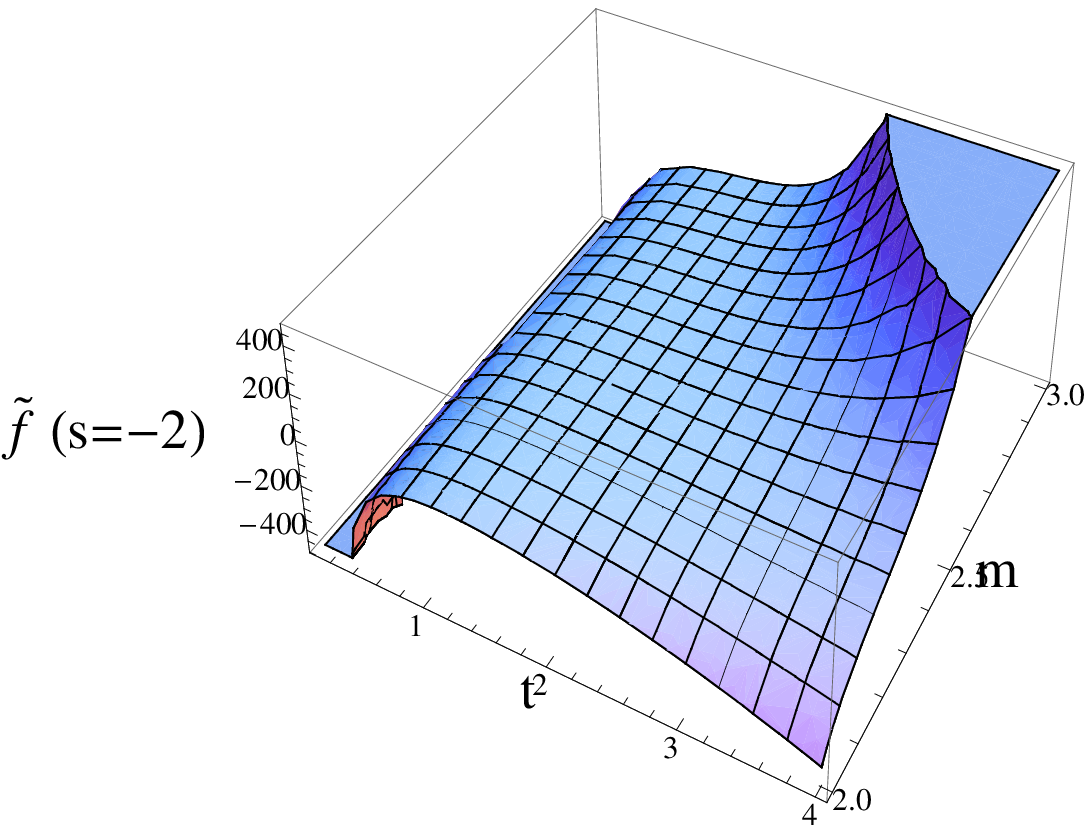}
\caption{Plot of reconstructed $f$ (Eq. (\ref{f-2})) from PDE taking
$s=-2$ and $m>1$. Also, $n=3,~C_3=0.2,~C_4=0.5$.} \label{M2restrict}
\hspace{0.5cm}\end{minipage}\hspace{0.5cm}
\begin{minipage}[b]{0.45\linewidth}
\centering\includegraphics[width=\textwidth]{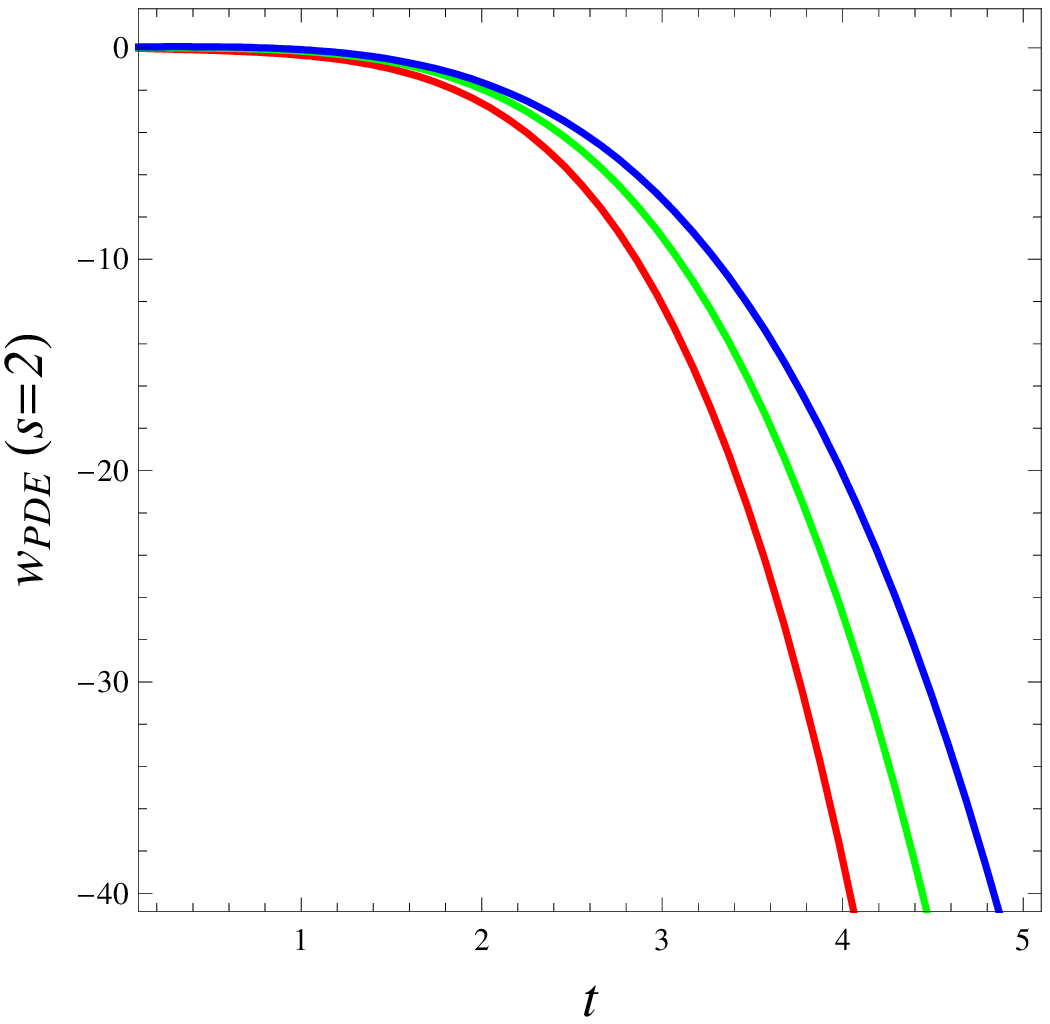}
\caption{Plot of the reconstructed EoS parameter (Eq. (\ref{w2}))
for $s=2$.} \label{M3restrict} \end{minipage}
\hspace{0.5cm}\begin{minipage}[b]{0.45\linewidth}
\centering\includegraphics[width=\textwidth]{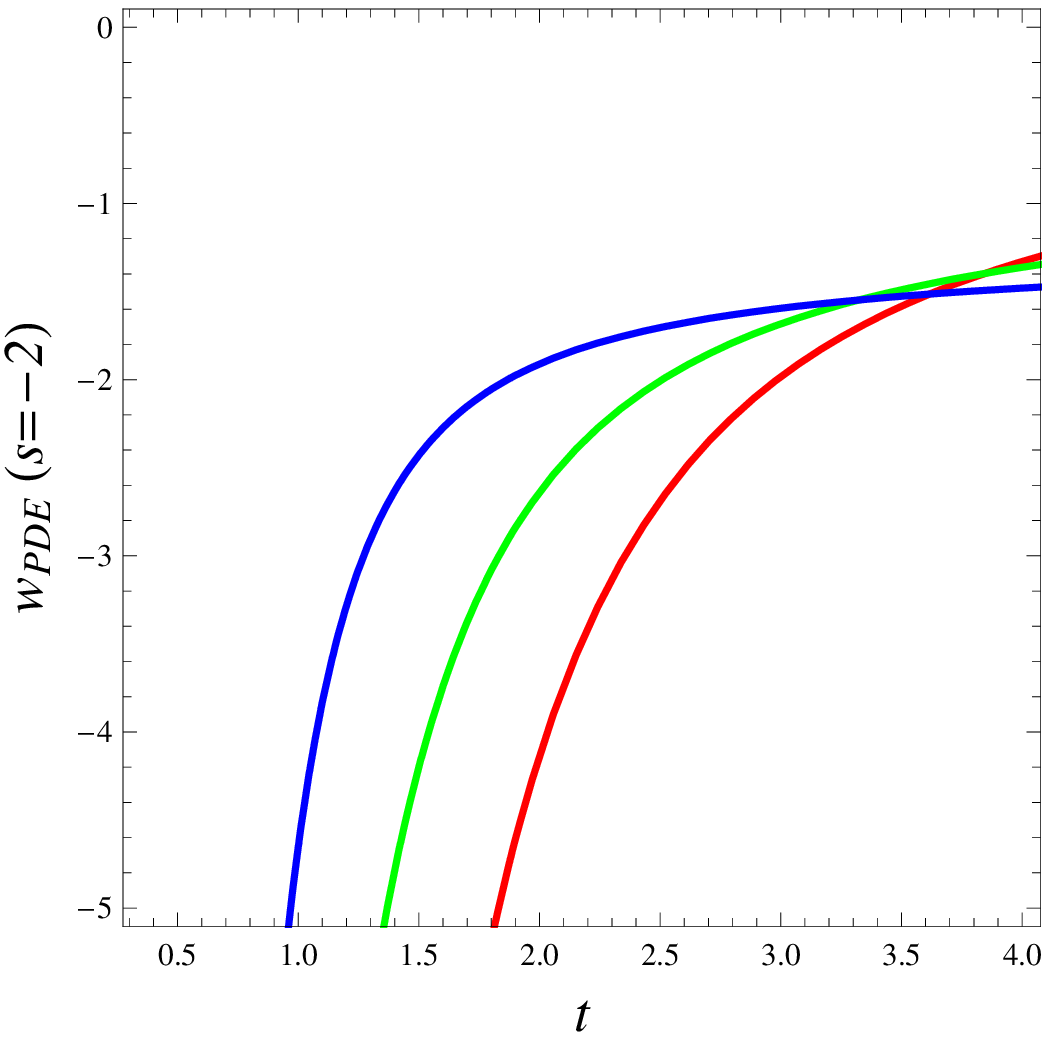}
\caption{Plot of the reconstructed EoS parameter (Eq. (\ref{w-2}))
for $s=-2$.} \label{cpleos} \end{minipage}\end{figure}

We also plot EoS parameter versus cosmic time corresponding to the
same values of PDE parameter for three different values of $m$ as
shown in FIG. \textbf{3} and \textbf{4}. It can be observed from
FIG. \textbf{3} that EoS parameter starts from dust-like matter,
passes the quintessence-like and vacuum DE era and then goes towards
phantom DE era. The $w_{DE}$ crosses phantom boundary at
$t\approx~0.5$ and transits from \emph{quintessence} to
\emph{phantom} i.e. behaves like \emph{quintom}. This plot also
represents that EoS parameter attains more reliable phantom era
which possesses the ability for prevention of BH formation. FIG.
\textbf{4} represents that EoS parameter remains in the phantom DE
era forever. The trajectories of EoS parameter corresponding to
different values of $m$ starts from higher phantom values and goes
towards less phantom values. Also, it is observed that it never
meets or crosses the $-1$ boundary and hence the EoS behaves like
\emph{phantom}. It is worthwhile to mention here that EoS parameter
corresponding to both cases of PDE parameter correspond to phantom
era of the universe which is a favorable sign to PDE conjecture.
However, the reconstructed model corresponing to $s=2$ is more
attractive as compare to $s=-2$ because of more aggressive phantom
era. However, taking into account the fact as stated in Wei (2012)
that for PDE $w_{DE}<-1$ always and it never crosses the phantom
divide $w_{DE}=-1$ in the whole cosmic history, it is interpreted
that EoS resulting from $s=-2$ is in complete agreement with it as
in this case $w_{DE}<-1$, $\rightarrow~-1$ and never crosses $-1$.
Moreover, $s=-2$ is a choice that is in agreement with the
prescription of Wei (2012), which states that if $\dot{H}<0$ (a
requirement for cosmic acceleration and holds for our choice of
scale factor) $s<0$. Hence, it is observed that the results stated
for PDE in Einstein gravity are in close agreement with those in the
framework of modified gravity under consideration.

\section{Reconstrction scheme for unification of
matter dominated and accelerated phases}
We consider that the Hubble rate $H$ is given by (Nojiri and Odintsov, 2006d,2006e)
\begin{equation}\label{hubbleassume}
H(t)=H_0+\frac{H_1}{t}.
\end{equation}
that leads to $a(t)=C_1 e^{H_0 t}t^{H_1}$ and due to this choice of Hubble parameter, the PDE takes the form
\begin{equation}\label{PDEnew}
\rho_{\Lambda}=3n^2 t^{-s}\left(H_0 t+H_1\right)^{s}
\end{equation}
When $t<<t_0$, in the early Universe and  $H(t)\sim \frac{H_1}{t}$, the Universe was filled with perfect fluid with
EOS parameter as
$w=−1+\frac{2}{3H_1}$. On the other hand, when $t>>t_0$ the Hubble parameter $H(t)$
is  constant $H \to H_0$ and the Universe seems to
de-Sitter . So, this form of $H(t)$ provides transitio from a
matter dominated to the accelerating phase. In Eq.
(\ref{modifiedrho}) we use (\ref{hubbleassume}) and we get the
following
\begin{eqnarray}\nonumber
\rho_{DE}&=&\frac{1}{2}\left[-f[t]+\left((H_1+H_0t)\left(-t^3
\left(10H_1^2(2+(-3+H_1)H_1)+H_0H_1(21+H_1(-73+40H_1))t\right.\right.\right.\right.\\\nonumber&+&
\left.\left.\left.\left.H_0^2\left(5-56H_1+60H_1^2\right)t^2+H_0^3
(-13+40H_1)t^3+10H_0^4t^4\right)f'[t]+(H_1+H_0t)\left(2H_1^2\right.\right.\right.\right.\\\nonumber&\times&
\left.\left.\left.\left.H_0t(-1+2H_0t)+H_1(-2+4H_0
t)\right)\left(12H_1\left(2H_1^2+H_0t(-1+2H_0t)+H_1(-2+4H_0
t)\right)\right)\right)\right.\right.\\\label{DEnew}&+&
\left.\left.t^4f''[t]\right)\left\{2H_1t^2\left(2H_1^2+H_0t(-1+2
H_0t)+H_1(-2+4H_0t)\right)^2\right\}^{-1}\right].
\end{eqnarray}
In the above as well as in the subsequent differential equations
$f'[t]~\textrm{and}~f''[t]$ denote the first and second order
derivatives of $f$ respectively with respect to $t$. If we consider
$\rho_{\Lambda}=\rho_{DE}$ as available in Eqs. (\ref{PDEnew}) and
(\ref{DEnew}), we get a differential equation that can not be solved
analytically for $f$. Hence, we solve it numerically to have $f$
graphically. Using the same approach as in the previous section, we
have reconstructed EoS parameter and showed it graphically. In FIGs.
\textbf{\ref{M1restrict}} and \textbf{\ref{M2restrict}}, we have
plotted $\tilde{f}$ for $s=2$ and $-2$, respectively. In case of
$s=2$, we have taken
$n=0.91,~H_1=2~(\textrm{red}),~2.2~(\textrm{green}),~2.3~(\textrm{blue})$
and $H_0=0.5$. For $s=-2$, we have taken
$n=4,~H_1=3~(\textrm{red}),~3.1~(\textrm{green}),~3.2~(\textrm{blue})$
and $H_0=0.89$.
\begin{figure}[ht] \begin{minipage}[b]{0.45\linewidth}
\centering\includegraphics[width=\textwidth]{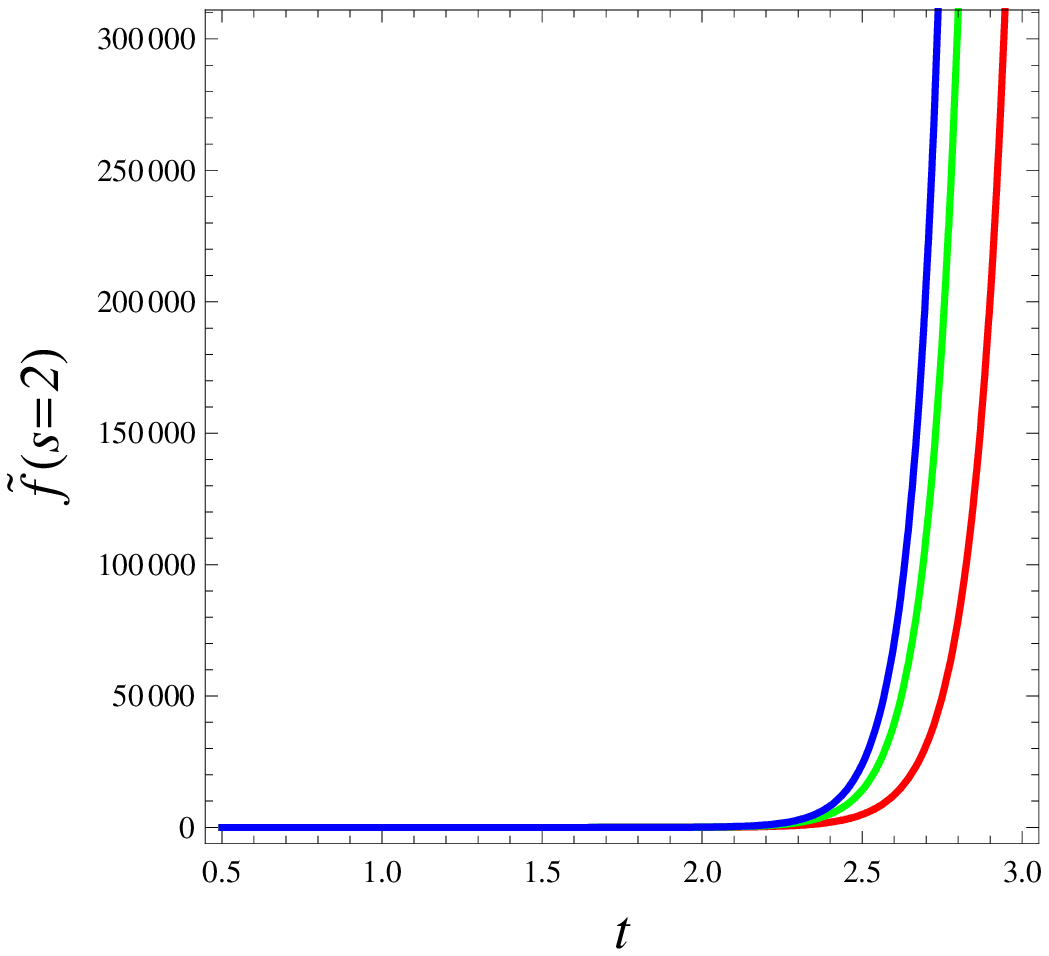}
\caption{Plot of reconstructed $\tilde{f}$ taking $s=2$ for
$H(t)=H_0+\frac{H_1}{t}$ .} \label{M1restrict} \end{minipage}
\hspace{0.5cm} \hspace{0.5cm}
\begin{minipage}[b]{0.45\linewidth}
\centering\includegraphics[width=\textwidth]{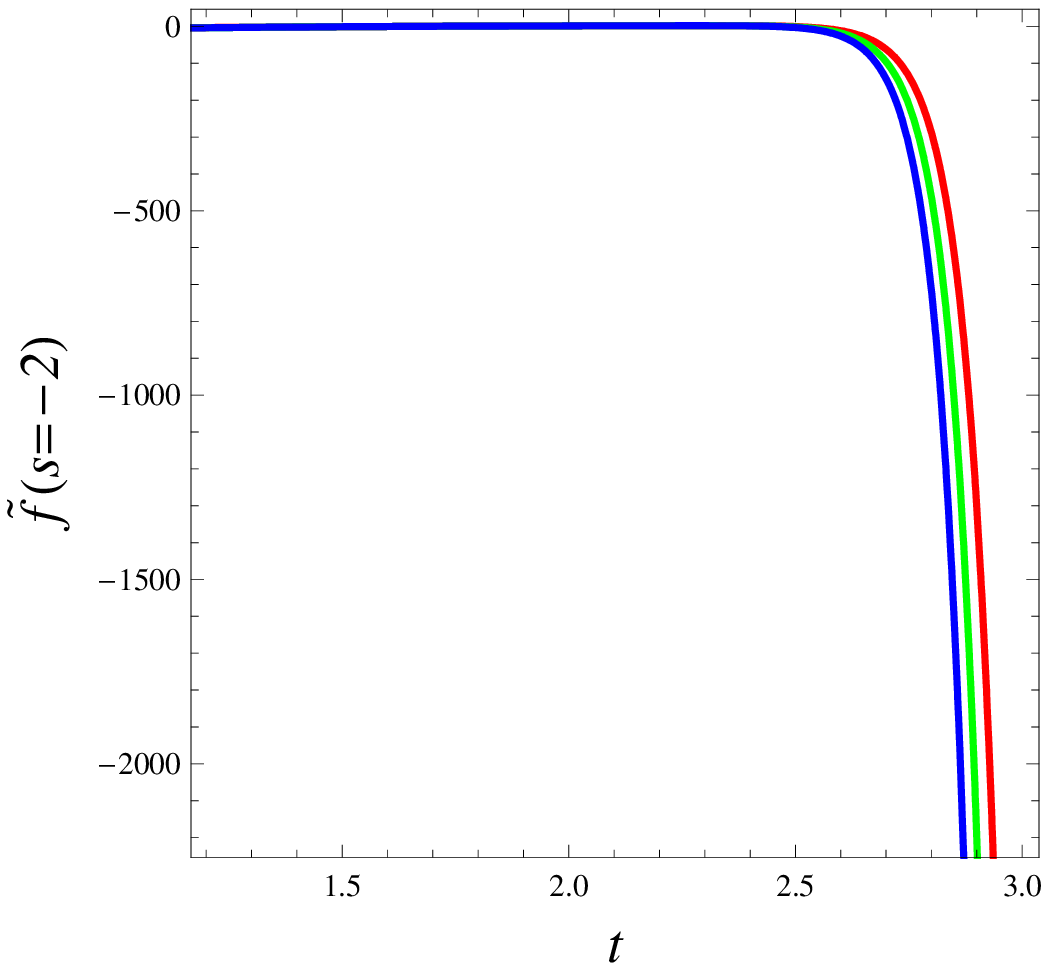}
\caption{Plot of reconstructed $\tilde{f}$ taking $s=-2$ for
$H(t)=H_0+\frac{H_1}{t}$.} \label{M2restrict}
\hspace{0.5cm}\end{minipage}\hspace{0.5cm}
\begin{minipage}[b]{0.45\linewidth}
\centering\includegraphics[width=\textwidth]{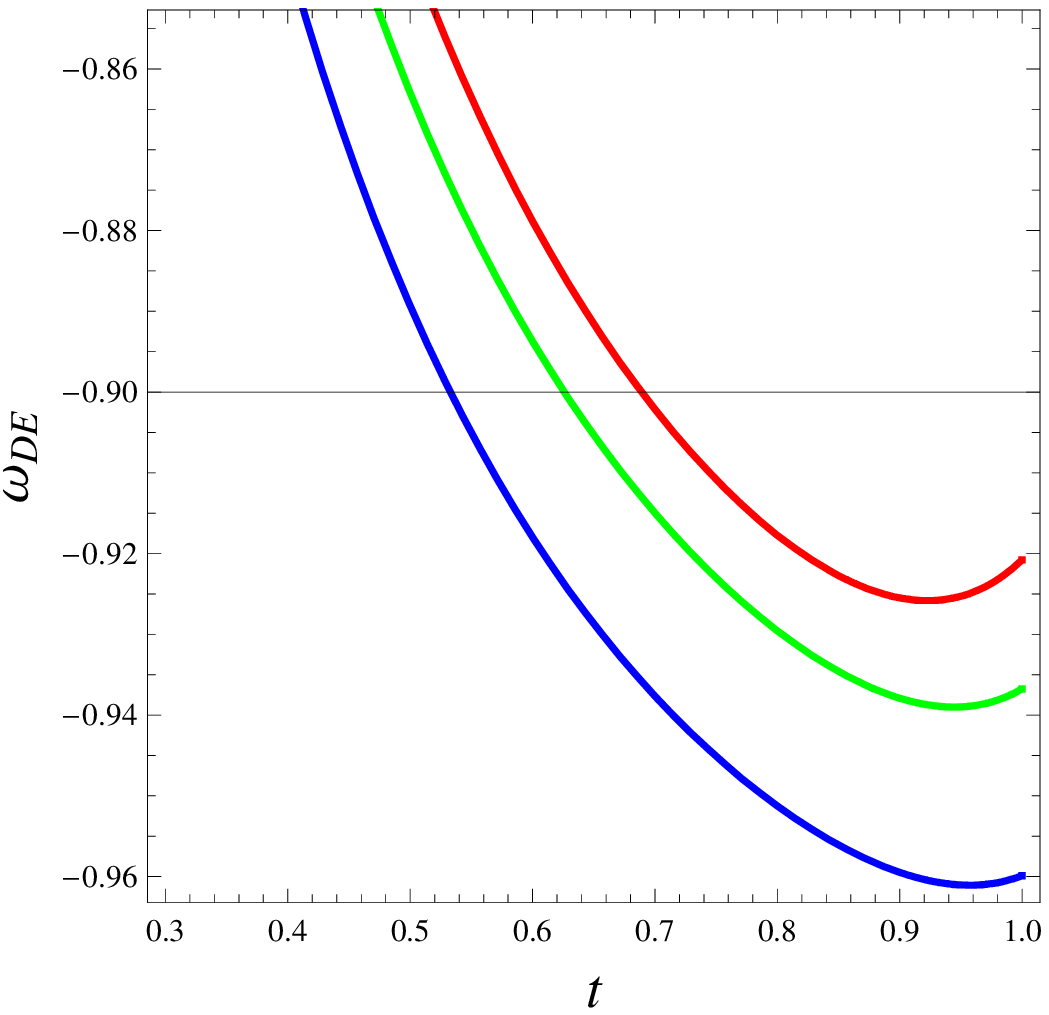}
\caption{Plot of the reconstructed EoS parameter for $s=2$.}
\label{M3restrict} \end{minipage}
\hspace{0.5cm}\begin{minipage}[b]{0.45\linewidth}
\centering\includegraphics[width=\textwidth]{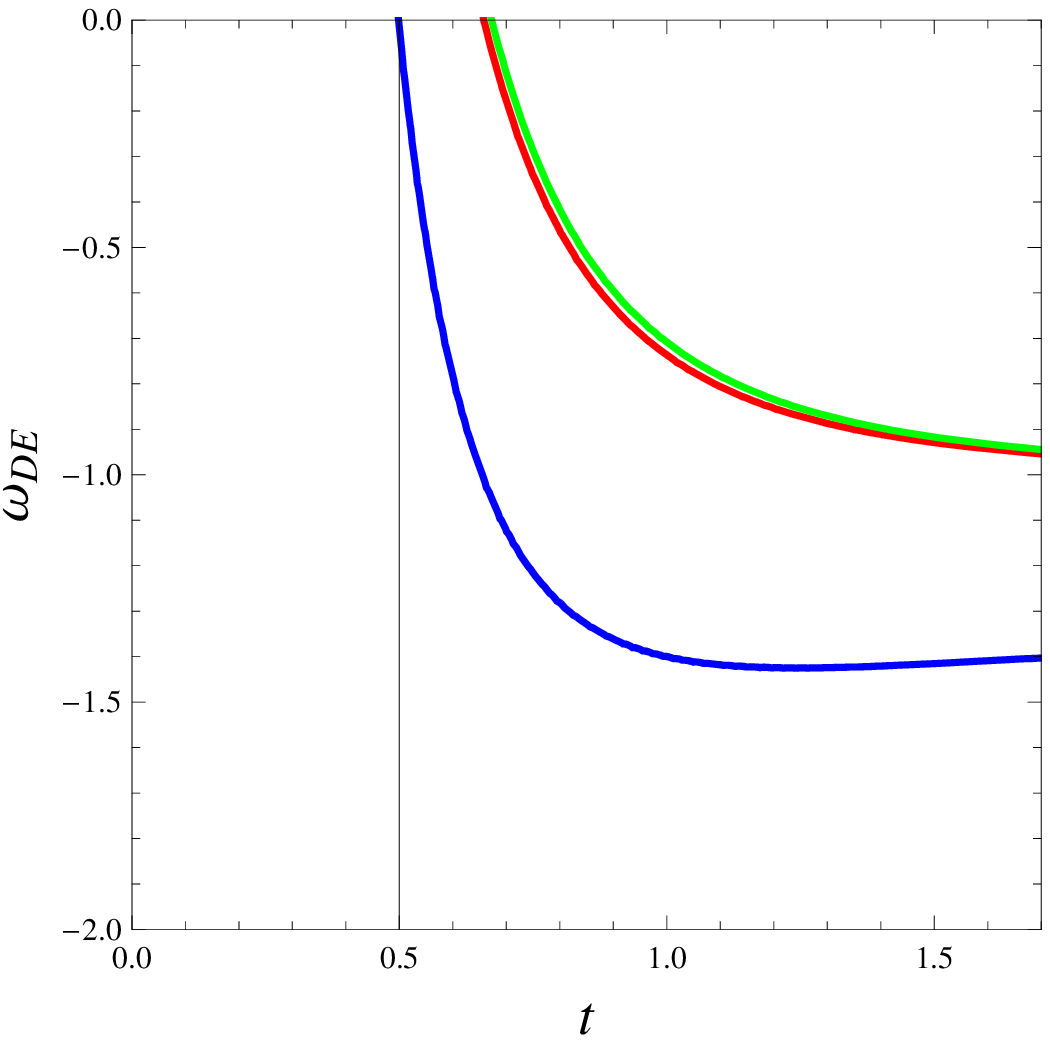}
\caption{Plot of the reconstructed EoS parameter for $s=-2$.}
\label{M4restrict} \end{minipage}\end{figure}

For $s=2$, we have observed increasing pattern of $\tilde{f}$ and in
case of $s=-2$, it is exhibiting decreasing pattern. We have plotted
the EoS parameters corresponding to $s=-2$ and $2$ in FIGs.
\textbf{\ref{M3restrict}} and \textbf{\ref{M4restrict}},
respectively. In case of $s=2$, the EoS parameter is $\geq-1$ and
hence it is behaving like quintessence. However, for $s=-2$, the EoS
parameter is crossing $-1$ boundary for $H_1=3.2$ and hence it is
behaving like quintom. Hence, it is understood that reconstructed
$f(T,T_G)$ through consideration of PDE can attain phantom era when
$s=-2$. This is consistent with the behavior of PDE that leads to
purely phantom era when considered in Einstein gravity with $s=-2$.
However, in $f(T,T_G)$ gravity, it can go beyond phantom for our
choice $H(t)=H_0+\frac{H_1}{t}$.
\subsection{On unification of inflation with DE in $f(T,T_G) $}
In this short subsection our aim is to realize $f(T,T_G)$ to unify the early inflationary epoch with the late time de-Sitter era. We mention here that
the
unification of inflation with DE in modified gravities firstly examined for $f(R)$ gravity. It was proposed in Nojiri-Odintsov model (Nojiri and Odintsov 2003), which was subsequently generalized to more realistic versions (Nojiri and Odintsov 2007d; Cognola et al. 2008). One important problem in early Universe is singularity and it investigated later (Nojiri and Odintsov 2008). Indeed it has been shown that there is a class of
non-singular exponential gravity to unify the  early- and late-time accelerated expansion of the Universe (Elizalde et al. 2011). For a review see (Nojiri and Odintsov 2011c).
\par
For our $f(T,T_G)$ case,
 if we consider the inflationary solution for  $H=\frac{H_1}{t}$ then we have
an exact solution of $f$ as
\begin{equation}
 \tilde{f}=\frac{1}{t^7}\left[\frac{48 n^2 (-t)^{7-s}}{8+(-14+s) s}-3 t^5+C_1t^{-\sqrt{41}} +C_2t^{\sqrt{41}} \right]
\end{equation}
This is an inflationary solution. We denote by $j=7-s>5,\ell=\frac{48 n^2 }{8+(-14+s) s}$, so we obtain:
 \begin{equation}
 \tilde{f}=\frac{1}{t^7}\left[\ell~t^j-3 t^5+C_1t^{-\sqrt{41}} +C_2t^{\sqrt{41}} \right]
\end{equation}
At inflationary (early) Universe, when $t<<t_0$,the dominant part of the $ \tilde{f}$ is written as follows:
\begin{equation}
 \tilde{f}\sim C_1t^{-(7+\sqrt{41})} .
\end{equation}
Since in this limit,
\begin{eqnarray}
T\sim\frac{6H_1^2}{t^2},\ \
T_G\sim\frac{24H_1^3(H_1-1)}{t^4}
\end{eqnarray}
So, the reconstructed $f(T,T_G)$ for inflationary era is written as the following:
\begin{equation}
f(T,T_G)= C_1\Big[-\frac{T_G}{24}\sqrt{\frac{6}{T}}\Big]^{7+\sqrt{41}} .
\end{equation}
So, $f(T,T_G)$ produces also the inflationary (phantom) solutions as well.\par
At late time,i.e. in the de-Sitter epoch when $H(t)\sim H_0$, we use of this fact that :
\begin{eqnarray}
T\sim 6H_0^2,\ \ T_G\sim 24 H_0^4  ,\ \ \dot{f}_{T_G}=\dot{T}_G f_{T_G,T_G}+\dot{T}f_{T_G,T}|_{H=H_0}=0
\end{eqnarray}
 so the reconstruction scheme gives us:
\begin{equation}
f(T,T_G)= \sqrt{T}F\left(\frac{T_G}{\sqrt{T}}\right)-T+\frac{6^{1-s/2}n^2m_p^{4-s}}{s-1}T^{s/2}.
\end{equation}
Where $F$ denotes an arbitrary function. The above $f(T,T_G)$  reproduces de-Sitter (late time) epoch.

\section{Reconstruction scheme for Intermediate Scale Factor}
Next, we consider the following scale factor (Barrow et al., 2006)
\begin{eqnarray}
&&a(t)=exp(At^m),\ \ 0<m<1.
\end{eqnarray}
The scale factor and Hubble parameter is suitably chosen so that it is consistent with the intermediate expansion:
\begin{eqnarray}
&&H(t)=Amt^{m-1}.
\end{eqnarray}
The scale factor is necessary to perform the analysis and therefore
working with a hypothetical scale factor may not be consistent with
the inflationary scenario. Hence we picked the intermediate scale
factor which is also consistent with astrophysical observations (Barrow et al., 2006).
Subsequently we have the following differential equation
\begin{eqnarray}\nonumber
3 n^2\left(\frac{t^{1-m}}{A
m}\right)^{-s}&=&\frac{1}{2}\left(6A^2m^2t^{-2+2m}-f[t]+\frac{t
f'[t]}{-1+m}+\frac{t\left(-1+m+Amt^m\right)f'[t]}{(-1+m)\left(-4+m
\left(3+4At^m\right)\right)}\right.\\\nonumber&+&
\left.\frac{t\left(\left(20+m^2\left(9+16At^m\right)-m\left(27+20A
t^m\right)\right)f'[t]-t\left(-4+m\left(3+4At^m\right)\right)
f''[t]\right)}{(-1+m)\left(-4+m\left(3+4At^m\right)\right)^2}\right)\\\label{diffinter}
\end{eqnarray}

\begin{figure}[ht] \begin{minipage}[b]{0.45\linewidth}
\centering\includegraphics[width=\textwidth]{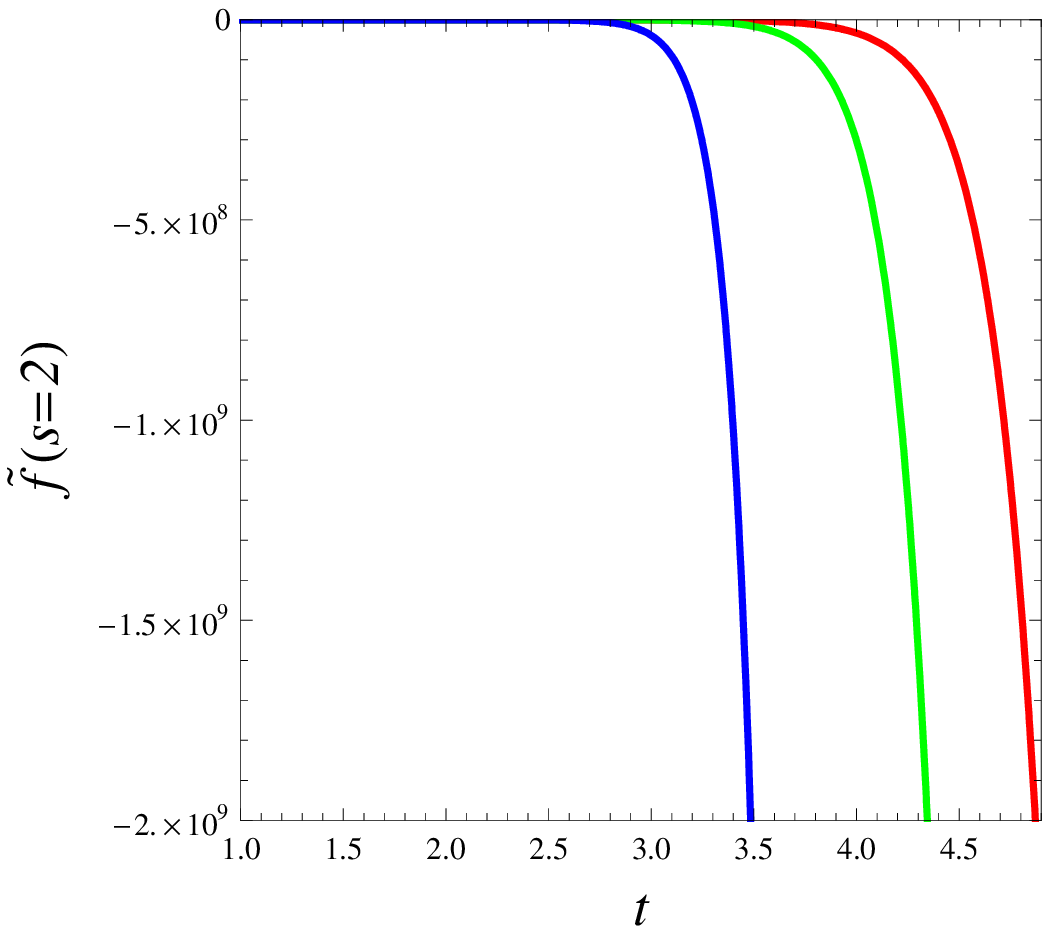}
\caption{Plot of reconstructed $\tilde{f}$ taking $s=2$ for
intermediate scale factor.} \label{M5restrict}
\end{minipage} \hspace{0.5cm} \hspace{0.5cm} \begin{minipage}[b]{0.45\linewidth}
\centering\includegraphics[width=\textwidth]{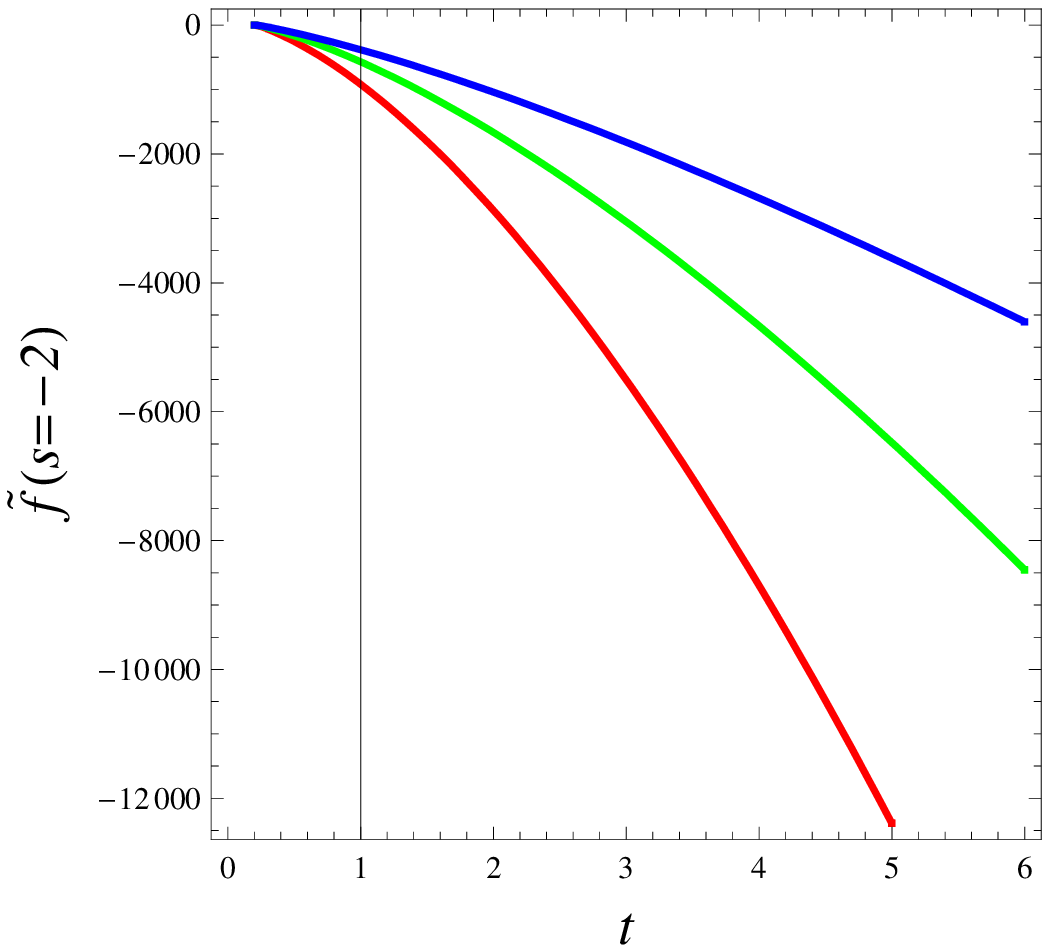}
\caption{Plot of reconstructed $\tilde{f}$ taking $s=-2$ for
intermediate scale factor.} \label{M6restrict}
\hspace{0.5cm}\end{minipage}\hspace{0.5cm}
\begin{minipage}[b]{0.45\linewidth}
\centering\includegraphics[width=\textwidth]{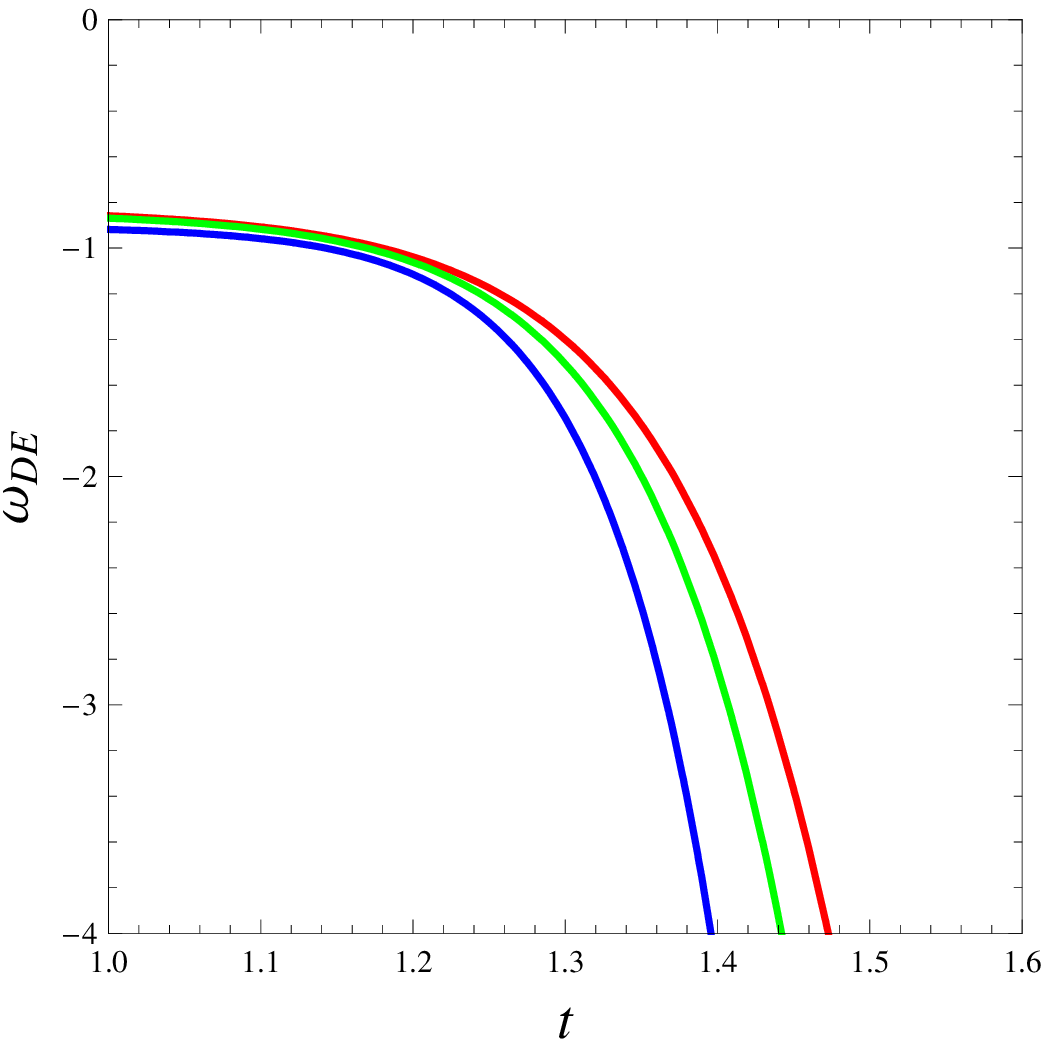}
\caption{Plot of the reconstructed EoS parameter for $s=2$ for
intermediate scale factor.} \label{M7restrict} \end{minipage}
\hspace{0.5cm}\begin{minipage}[b]{0.45\linewidth}
\centering\includegraphics[width=\textwidth]{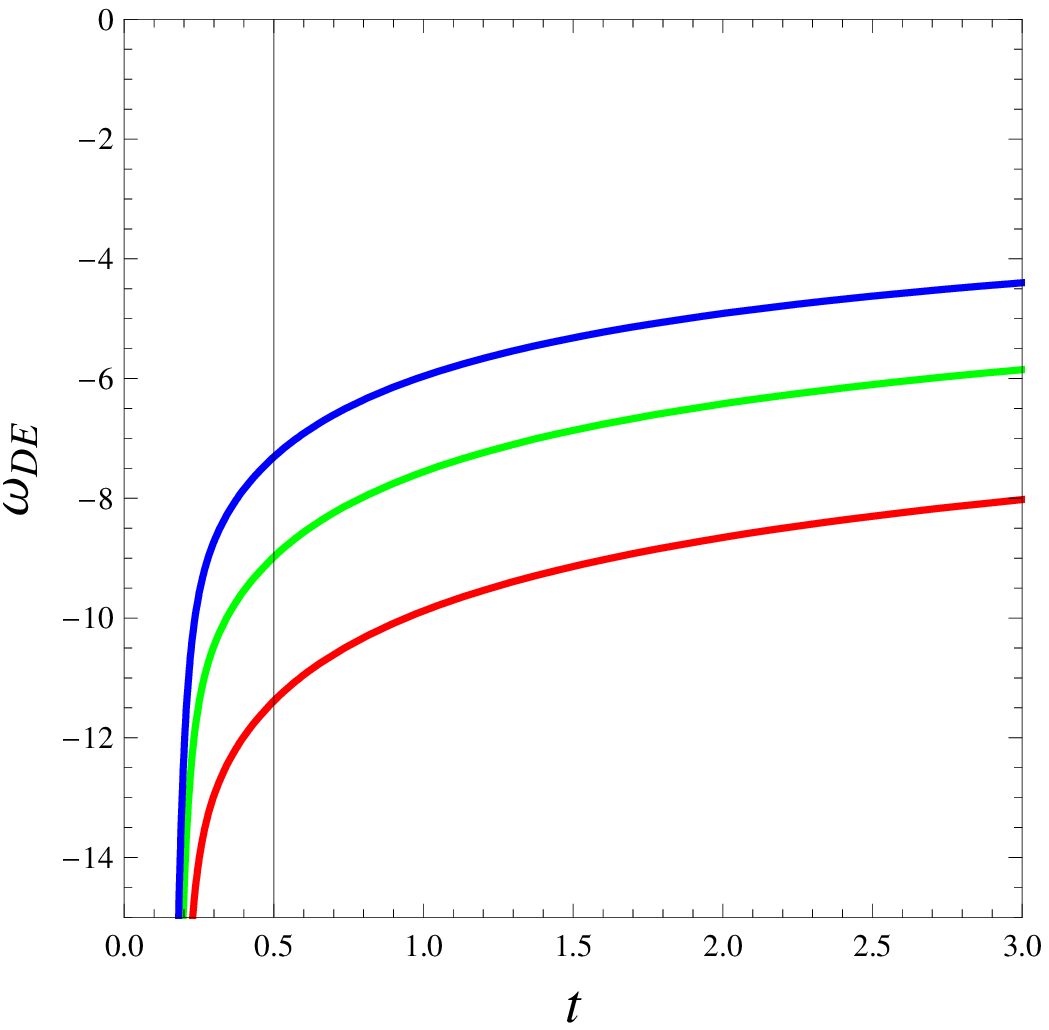}
\caption{Plot of the reconstructed EoS parameter for $s=-2$ for
intermediate scale factor.} \label{M8restrict}
\end{minipage}\end{figure}

Solving Eq. (\ref{diffinter}) numerically and plotting reconstructed
$\tilde{f}$ in FIGs. \textbf{\ref{M5restrict}} and
\textbf{\ref{M6restrict}} for $s=-2$ and $s=2$. We observe that for
$s=2$ as well as $s=-2$, the reconstructed $f(T,T_G)$ model is
displaying decreasing pattern. It is noted from FIG.
\textbf{\ref{M7restrict}} (for $s=2$) that the reconstructed
$w_{DE}$ crosses phantom boundary at $t\approx1.15$ and hence
behaves like quintessence. However, in Fig.
\textbf{\ref{M8restrict}} (for $s=-2$), we observe that the EoS
parameter $w_{DE}<-1$ which indicates an aggressively phantom-like
behavior. Hence, for intermediate scale factor, the PDE in modified
gravity $f(T,T_G)$ pertains to phantom era of the universe and hence
it is consistent with the behavior of PDE in Einstein gravity for
$s=-2$. It may be noted (for $s=2$ case) that we have taken
$n=0.9,~A=10$ and red, green and blue lines correspond to
$m=0.33,~0.35,~0.40$ respectively. On the other hand, for $s=-2$, we
have taken $n=4,~A=0.6$ and red, green and blue lines correspond to
$m=0.20,~0.25,~0.30$ respectively.

\section{Reconstruction scheme For Bouncing Scale Factor}
Inflation is a solution for flatness problem in big-bang cosmology. Bounncing scenario predicts a transitionary inflationary Universe,in which the Universe evoles from a contracting epoch $(H<0)$ to an expanding epoch $(H>0)$. It means the scale factor $a(t)$ reaches a local minima. So the cosmological solution is non-singular. In GB gravity,bouncing solutions widely studied in literature (Bamba et al. 2014a,2104b; Odintsov et al. 2014; Makarenko et al. 2014; Amoros et al. 2013; Nojiri et al. 2003; Lidesy et al. 2002a,2002b).

This scale factor takes the following form (Myrzakulov and Sebastiani, 2014)
\begin{equation}
a(t)=a_0+\alpha(t-t_0)^{2n}\,,\quad H (t)=\frac{2n\alpha(t-t_0)^{2n-1}}{a_0+\alpha(t-t_0)^{2n}}\,,\quad n=1,2,3...
\label{pow}
\end{equation}
where $a_0$, $\alpha$ are positive (dimensional) constants and $n$
is a positive natural number. The time of the bounce is fixed at
$t=t_0$. When $t<t_0$, the scale factor decreases and we have a
contraction with negative Hubble parameter. At $t=t_0$, we have the
bounce, such that $a(t=t_0)=a_0$, and when $t>t_0$ the scale factor
increases and the universe expands with positive Hubble parameter.
It should be mentioned that for sake of simplicity (without any loss
of generality) we have taken $n$ in the power law form as well as in
the PDE density (Eq. \ref{pde}). For this choice of scale factor, we
get the following differential equation
\begin{figure}[ht] \begin{minipage}[b]{0.45\linewidth}
\centering\includegraphics[width=\textwidth]{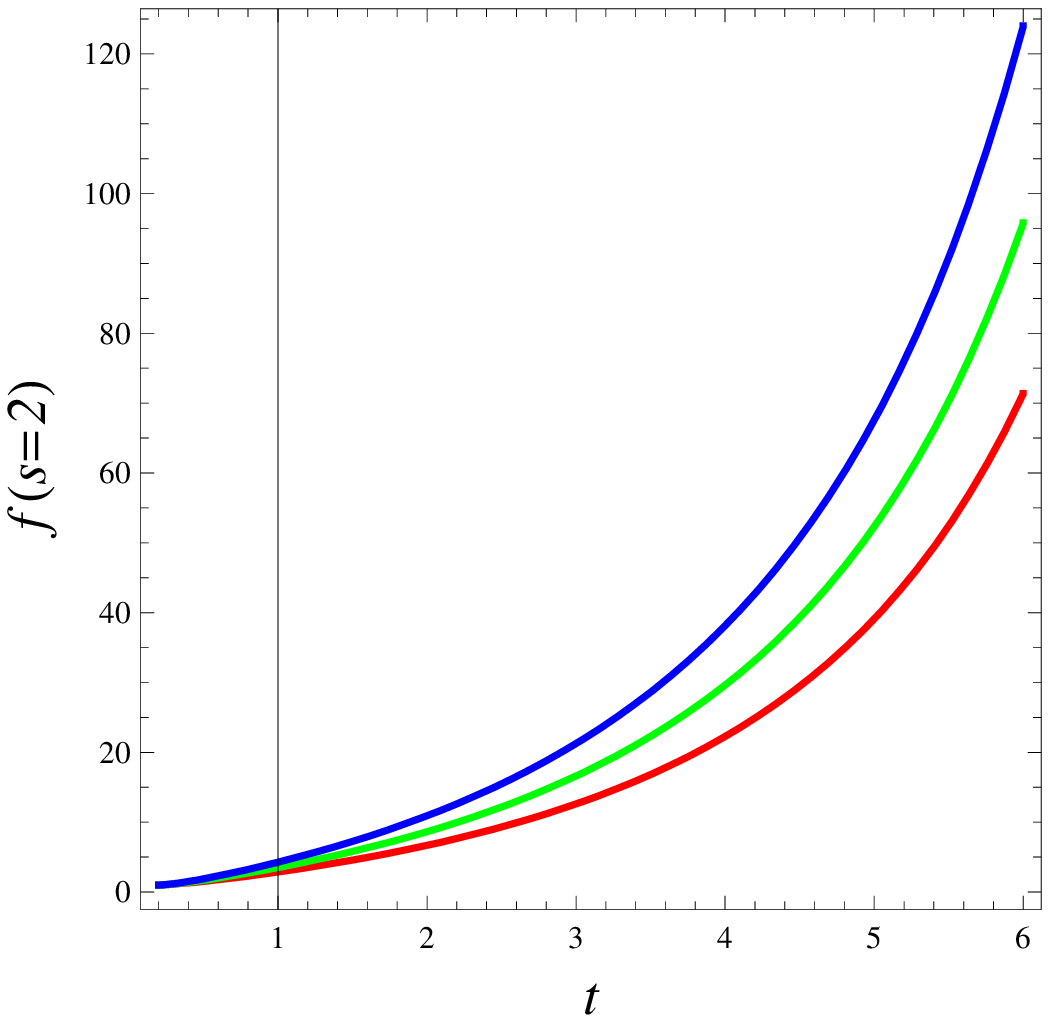}
\caption{Plot of reconstructed $\tilde{f}$ taking $s=2$ for bounce
with power-law scale factor.} \label{M9restrict} \end{minipage}
\hspace{0.5cm} \hspace{0.5cm} \begin{minipage}[b]{0.45\linewidth}
\centering\includegraphics[width=\textwidth]{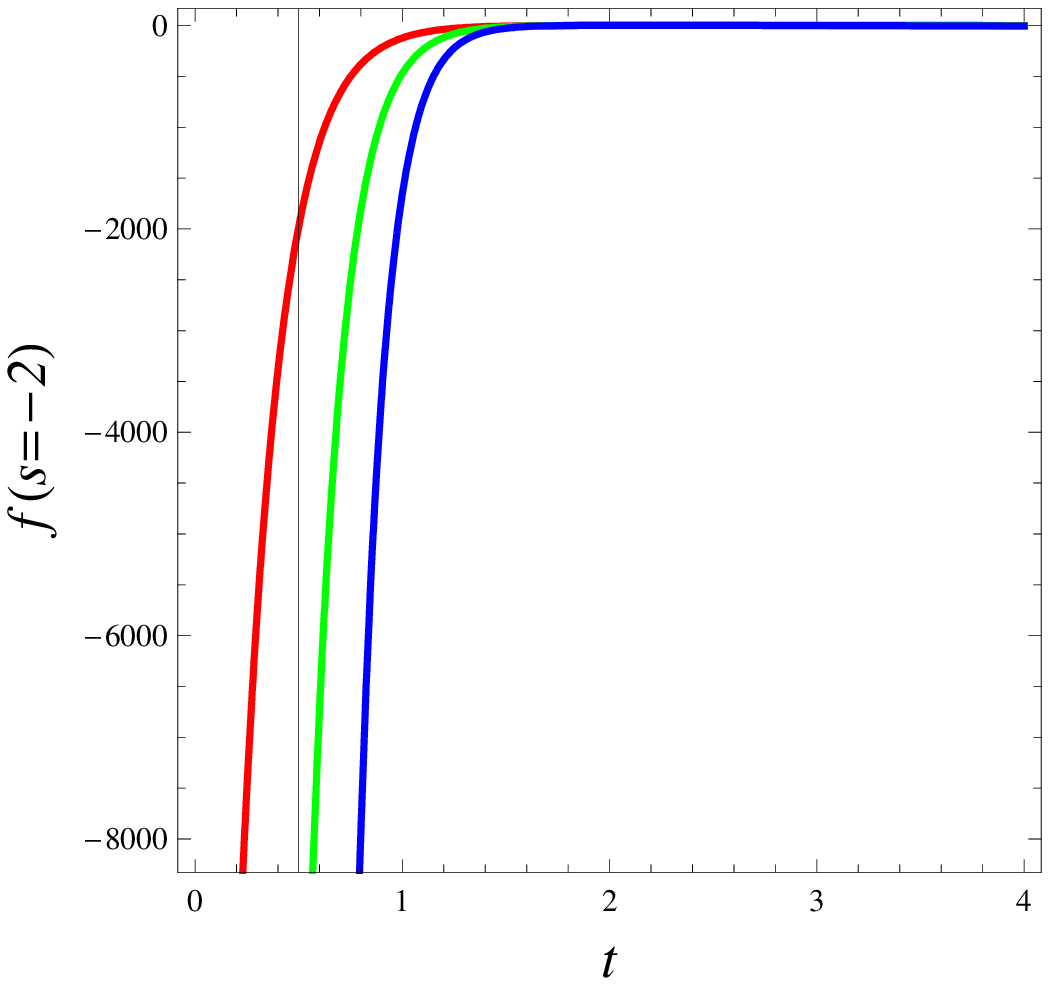}
\caption{Plot of reconstructed $\tilde{f}$ taking $s=-2$ for bounce
with power-law scale factor.} \label{M10restrict}
\hspace{0.5cm}\end{minipage}\hspace{0.5cm}
\begin{minipage}[b]{0.45\linewidth}
\centering\includegraphics[width=\textwidth]{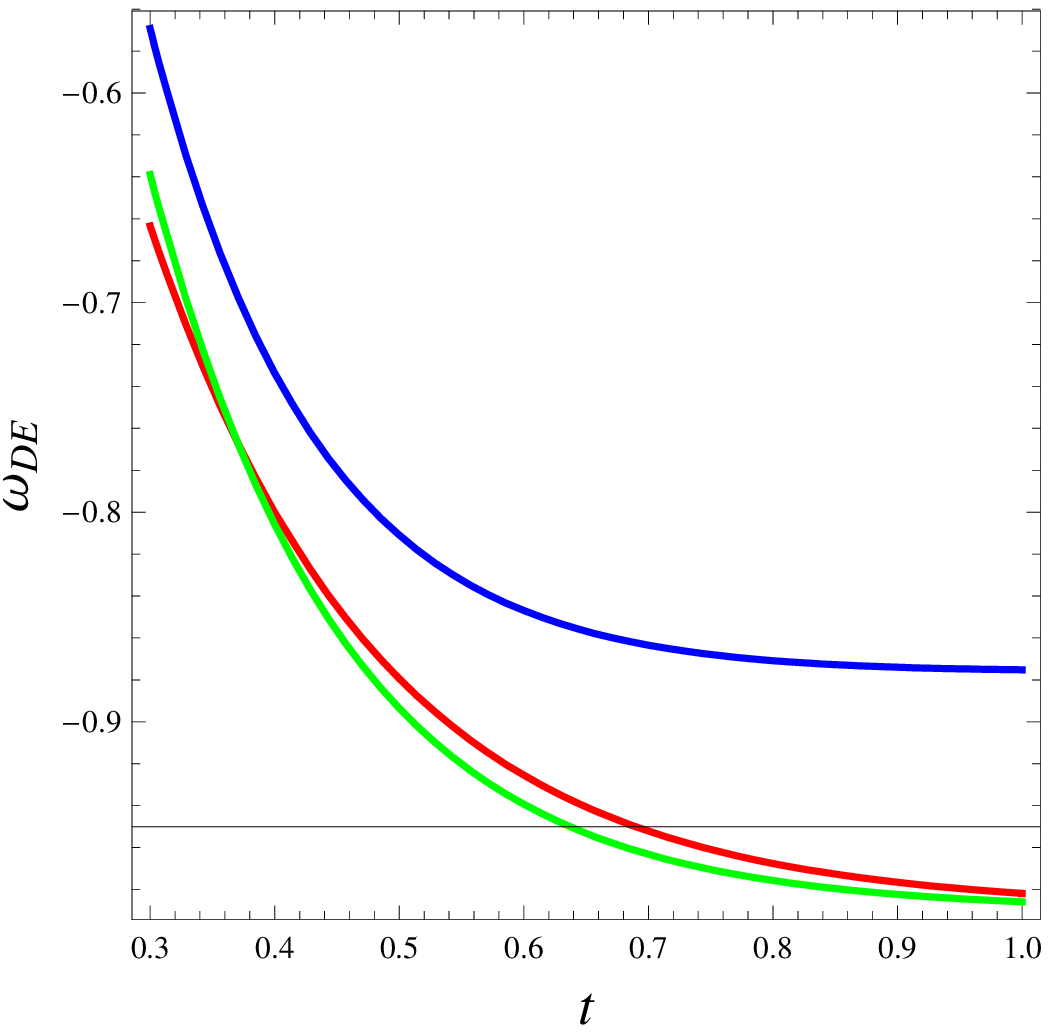}
\caption{Plot of the reconstructed EoS parameter for $s=2$ for
bounce with power-law scale factor.} \label{M11restrict}
\end{minipage} \hspace{0.5cm}\begin{minipage}[b]{0.45\linewidth}
\centering\includegraphics[width=\textwidth]{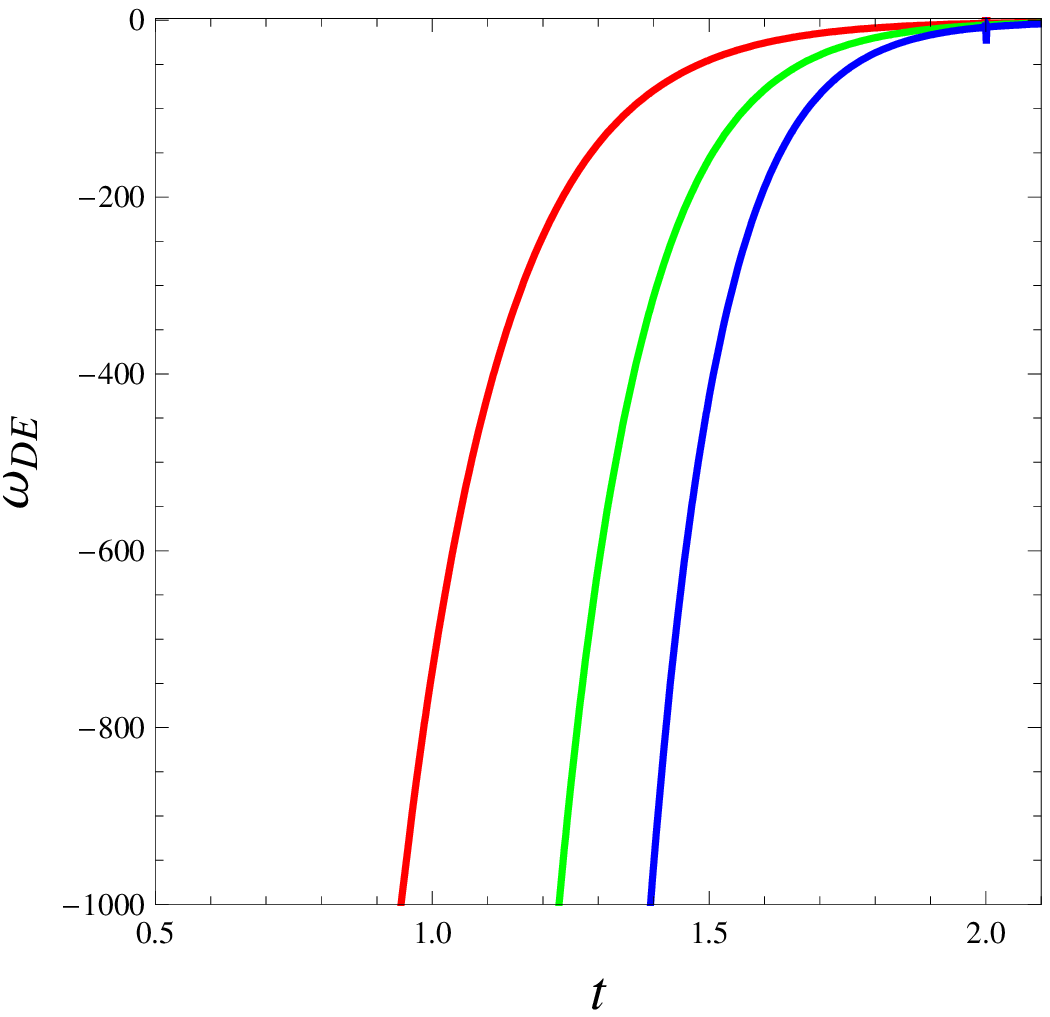}
\caption{Plot of the reconstructed EoS parameter for $s=-2$ for
bounce with power-law scale factor.} \label{M12restrict}
\end{minipage}\end{figure}

\begin{eqnarray}\nonumber
&&\frac{1}{4}\left[\frac{48 n^2
(t-t_0)^{-2+4n}\alpha^2}{\left(a_0+(t-t_0)^{2
n}\alpha\right)^2}+\frac{(t-t_0)\left(a_0+(t-t_0)^{2
n}\alpha\right)f'[t]}{a_0(-2+3n)-2(t-t_0)^{2n}\alpha}-\frac{2(t-t_0)\left(a_0+(t-t_0)^{2
n}\alpha\right)f'[t]}{a_0-2a_0n+(t-t_0)^{2
n}\alpha}\right.\\\nonumber&-& \left.2
f[t]\left((t-t_0)\left(\left(a_0^2(-2+3n)(-5+6n)-a_0
(-20+3n(9+2n))(t-t_0)^{2n}\alpha+10(t-t_0)^{4n}\alpha^2\right)\right.\right.\right.\\\nonumber&\times&
\left.\left.\left.f'[t]-(t-t_0)
\left(a_0(-2+3n)-2(t-t_0)^{2n}\alpha\right)\left(a_0+(t-t_0)^{2
n}\alpha\right)f''[t]\right)\right)((-1+2n)\left(a_0(2-3n)\right.\right.\\\label{bounce}&\times&
\left.\left.+2(t-t_0)^{2n}\alpha
\right)^2)^{-1}\right]=32^sn^2\left(\frac{(t-t_0)^{1-2n}\left(a_0+(t-t_0)^{2n}\alpha
\right)}{n\alpha}\right)^{-s}.
\end{eqnarray}
Solving Eq. (\ref{bounce}) numerically and plotting reconstructed
$\tilde{f}$ in FIGs. \textbf{\ref{M9restrict}} and
\textbf{\ref{M10restrict}}, we observe that the reconstructed
$f(T,T_G)$ is displaying increasing pattern for $s=2$. However, for
$s=-2$, the reconstructed $f(T,T_G)$ is displaying increasing
pattern and is tending to $0$ at late stage of the universe. It is
also noted from FIG. \textbf{\ref{M11restrict}} (for $s=2$) that the
reconstructed $w_{DE}\rightarrow-1$ from $w_{DE}>-1$. However, it
never crosses phantom boundary and hence behaves like quintessence
throughout. However, in FIG. \textbf{\ref{M12restrict}}, it can be
observed that for $s=-2$ the EoS parameter $w_{DE}<-1$ and this
indicates aggressive phantom-like behavior. Hence, with $s=-2$ for
bounce with power-law scale factor, PDE in modified gravity
$f(T,T_G)$ pertains to phantom era of the universe and hence it is
consistent with the behavior of PDE in Einstein gravity for $s=-2$.
It may be noted that for both of the cases, we have taken
$n=2,~A=10$ and red, green and blue lines correspond to
$m=0.33,~0.35,~0.40$ respectively. On the other hand, for $s=-2$ we
have taken $n=4,~a_0=10.5,~\alpha=10.1$ and red, green and blue
lines correspond to $n=6,~7,~8$ respectively.

\section{Reconstruction through a semi analytic form of $f$}
We assume that $f(T,T_G)$ realizes the form
\begin{equation}
f(T,T_G)	\equiv f=b_0+b_1~t+b_2~t^2+b_3~t^3+...
\end{equation}
For this choice of $f$, Eq. (\ref{modifiedrho}) takes the form
\begin{equation}
\begin{array}{c}
\rho_{DE}=\frac{1}{2} \left[-b_0-b_1 t-b_2 t^2-b_3 t^3+6 H^2+\frac{(b_1+t (2 b_2+3 b_3 t)) H}{H'}+\frac{(b_1+t
(2 b_2+3 b_3 t)) H \left(H^2+H'\right)}{2 H' \left(2 H^2+H'\right)+H H''}-\right.\\
\left.\frac{H \left(48 (b_2+3 b_3
t) H \left(2 H' \left(2 H^2+H'\right)+H H''\right)-
24 (b_1+t (2 b_2+3 b_3 t)) \left(2 H'^3+4 H^3 H''+6
H H' H''+H^2 \left(12 H'^2+H^3\right)\right)\right)}{24 \left(2 H' \left(2 H^2+H'\right)+H H''\right)^2}\right]
\end{array}
\end{equation}
\begin{figure}[ht] \begin{minipage}[b]{0.45\linewidth}
\centering\includegraphics[width=\textwidth]{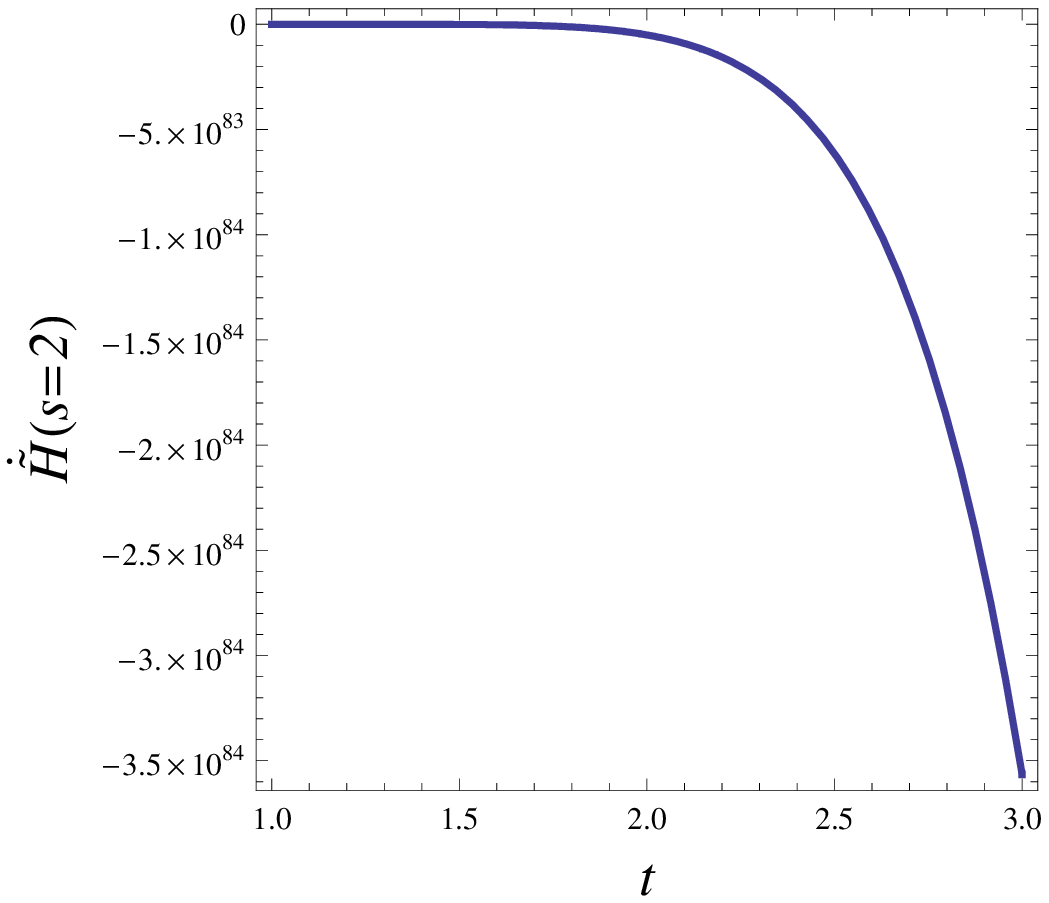}
\caption{Plot of reconstructed $\tilde{H}$ for $s=2$.}
\label{M17restrict} \end{minipage} \hspace{0.5cm} \hspace{0.5cm}
\begin{minipage}[b]{0.45\linewidth}
\centering\includegraphics[width=\textwidth]{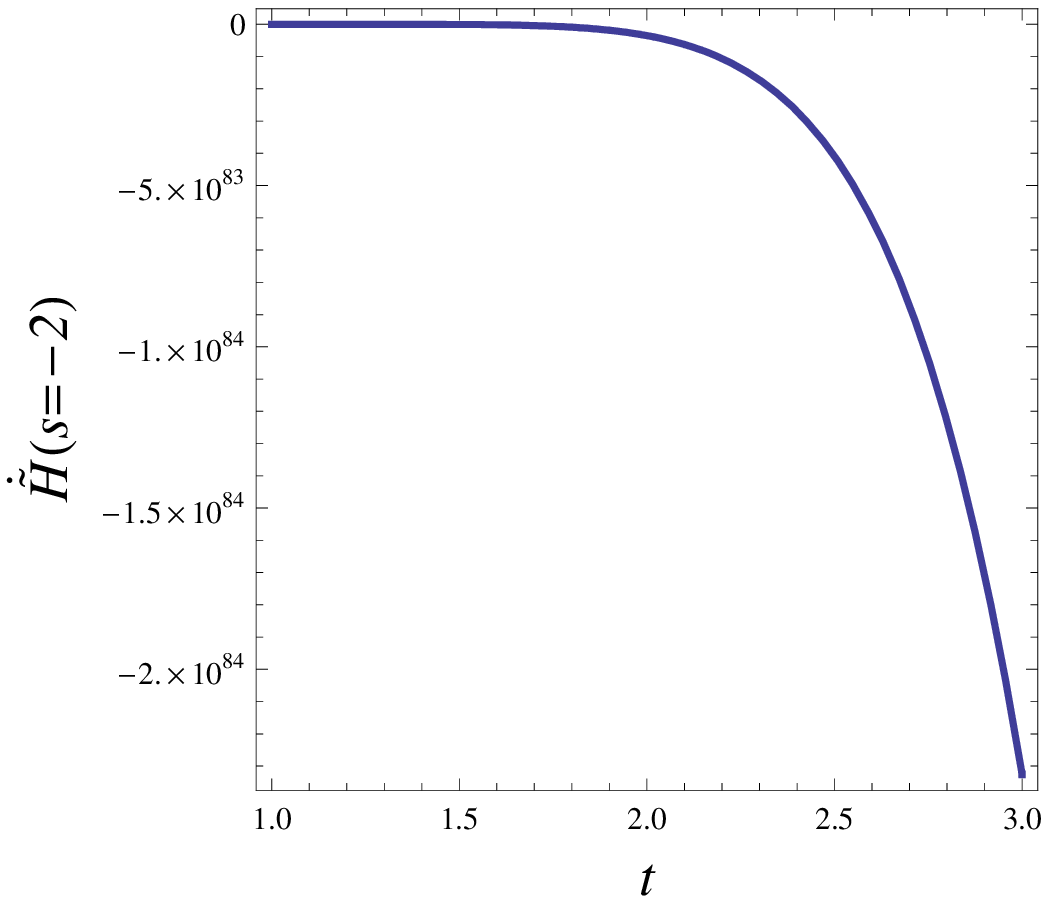}
\caption{Plot of reconstructed $\tilde{H}$ for $s=-2$.}
\label{M18restrict} \hspace{0.5cm}\end{minipage}\hspace{0.5cm}
\begin{minipage}[b]{0.45\linewidth}
\centering\includegraphics[width=\textwidth]{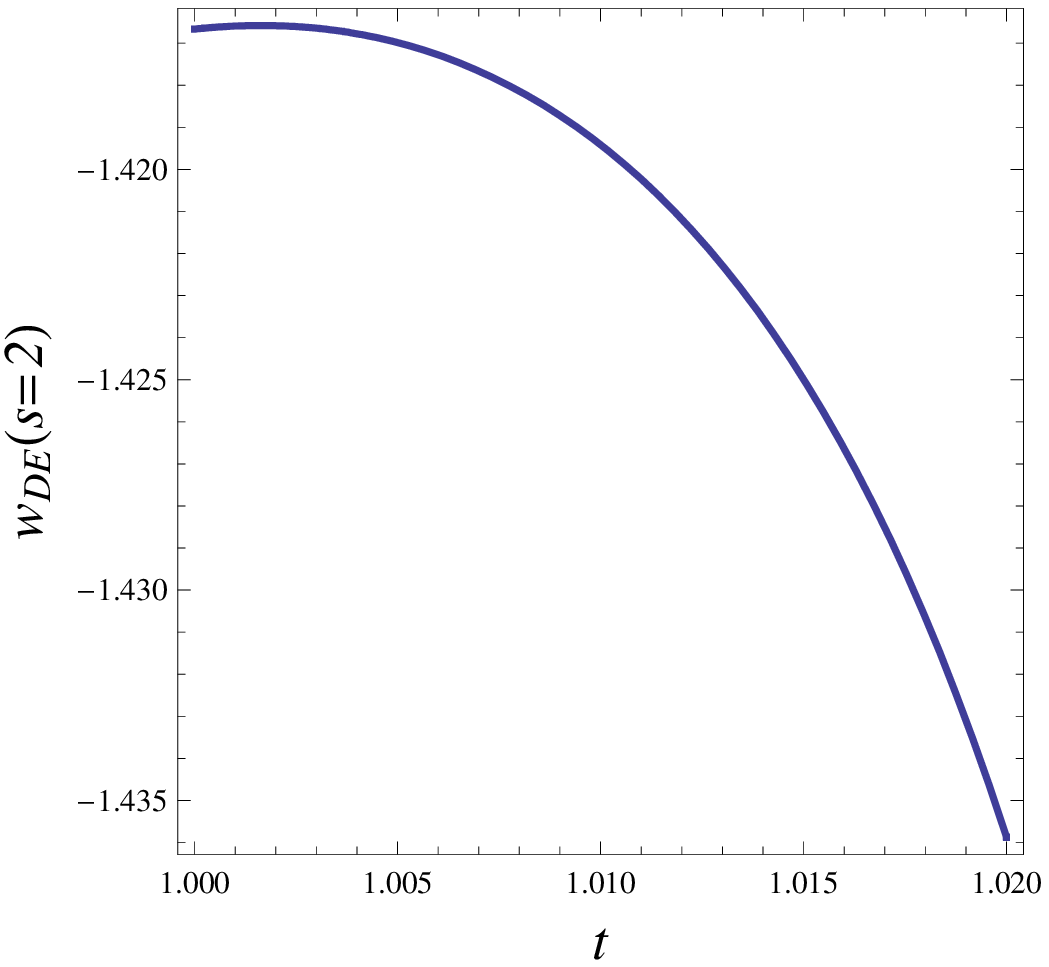}
\caption{Plot of the reconstructed EoS parameter for $s=2$.} \label{M19restrict} \end{minipage}
\hspace{0.5cm}\begin{minipage}[b]{0.45\linewidth}
\centering\includegraphics[width=\textwidth]{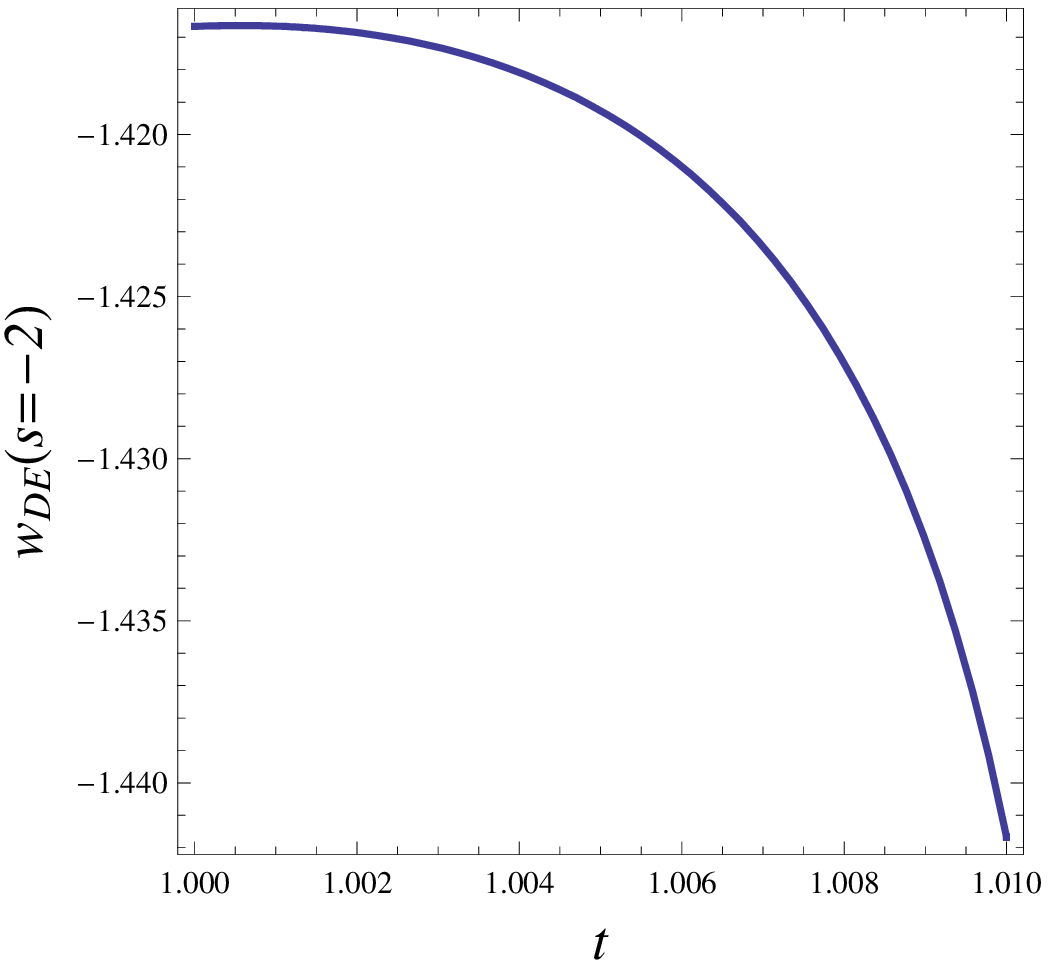}
\caption{Plot of the reconstructed EoS parameter for $s=-2$.} \label{M20restrict} \end{minipage}\end{figure}

which, on setting equal to $\rho_{\Lambda}$ gives rise to a
differential equation on $H$ that is solve numerically to generate
the reconstruction of $H$ and subsequently EoS parameter
$w_{DE}=-1-\frac{2\dot{H}}{3 H^2}$ that are plotted in Figs.
\ref{M17restrict}-\ref{M20restrict}. We have set the coefficients of
the polynomial as $b_0=0.8,~b_1=0.6,~b_2=0.5,~b_3=0.2$. In this
reconstruction through an assumed polynomial form of $f$, the Hubble
parameter $H$ has been reconstructed and the reconstructed $\tilde{H}$ has been examined for its first time derivative and it is observed that $\dot{\tilde{H}}<0$ throughout for $s=\pm~2$ and
$w_{DE}<-1$ for $s=\pm~2$. Negative time derivative of Hubble
parameter is consistent with the accelerated expansion of the
universe and the aggressive phantom-like behavior of $w_{DE}$ is
consistent with the basic property of pilgrim dark energy.
\section{Concluding remarks}
Pilgrim dark
energy (PDE)  model is studied in this paper and Hubble horizon has been used as
 an IR cutoff.  The basic assumption of this model is that phantom acceleration prevents the formation of the BH. The said PDE is considered in a modified gravity $f(T,
T_G)$, which has been constructed by Kofinas and Saridakis (2014) on
the basis of $T$ (old quadratic torsion scalar) and $T_G$ (new
quartic torsion scalar $T_G$ that is the teleparallel equivalent of
the Gauss-Bonnet term). We have compiled our work in two phases:
Firstly, we have assumed different scale factors such as $a(t)=a_0
t^m$, $H=H_0+\frac{H_1}{t}$, $a(t)=exp(At^m)$ and
$a(t)=a_0+\alpha(t-t_0)^{2n}$. We have reconstructed $f$ and
subsequently $w_{DE}$ in this scenario. Secondly, we have assumed
analytic function such $f=b_0+b_1~t+b_2~t^2+b_3~t^3$ and
reconstructed Hubble parameter and $w_{DE}$ without any choice of
scale factor.

Throughout the study, we have considered $s=-2$ and $s=2$,
separately. We have observed that $s=-2$, as described in PDE (Wei,
2012), seems more realistic choice for $s$ than $s=2$ and this
outcome of the present reconstruction work is consistent with Wei
(2012). Moreover, it has been observed that the reconstructed
$w_{DE}$, irrespective of choices of scale factor or a choice of
$f$, exhibit a more aggressive phantom-like behavior for $s=-2$ than
$s=2$. This result also matches the study of Wei (2012). Hence, it
is finally concluded that PDE, when considered in $f(T,T_G)$ gravity
is capable of attaining the phantom phase of the universe.
\vspace{0.5cm}

{\bf Acknowledgment}

\vspace{0.5cm}

Sincere thanks are due to the anonymous reviewer for constructive suggestions. The first author (SC) wishes to acknowledge the financial support
from Department of Science and Technology, Govt. of India under
Project Grant no. SR/FTP/PS-167/2011.

\section{References}

   Abazajian, K., et al.: Astron. J. 128, 502 (2004)\\
Abazajian, K., et al.: Astron. J. 129, 1755 (2005)\\
Allen, S.W., et al.: Mon. Not. R. Astron. Soc. 353, 457 (2004)\\
Amorós, J.~, et al.:
  Phys.\ Rev.\ D {\bf 87}, no. 10, 104037
  : arXiv:1305.2344 [gr-qc](2013)\\
Aslam, A., et al.:
  Can.\ J.\ Phys.\  {\bf 91}  93
  : arXiv:1212.6022 [astro-ph.CO] (2013)\\
Babichev, E. et al.: Phys. Rev. Lett.
\textbf{93}, 021102 (2004)\\
Babichev, E. et al.: Phys. Rev. D \textbf{78}, 104027 (2008)\\
Bamba, K. et al.: Astrophys.\ Space Sci.\ {\bf 341}, 155 (2012)\\
Bamba, K.,Phys.\ Rev.\ D {\bf 85}  104036: arXiv:1202.4057 [gr-qc](2012)\\
Bamba, K., et al.:  Astrophys.\ Space Sci.\  {\bf 344}, 259  : arXiv:1202.6114 [physics.gen-ph] (2013)\\
Bamba, K., et al.:Phys.\ Lett.\ B {\bf 725} 368  : arXiv:1304.6191 [gr-qc] (2013)\\
Bamba, K., Odintsov, S. D., Sáez-Gómez, D.:Phys.\ Rev.\ D {\bf 88}  084042  : arXiv:1308.5789 [gr-qc](2013) \\
Bamba, K., Nojiri, S., Odintsov, S. D.:Phys.\ Lett.\ B {\bf 731} 257: arXiv:1401.7378 [gr-qc] (2014)\\
Bamba, K., Odintsov, S. D.:arXiv:1402.7114 [hep-th] (2014)\\
Bamba, K., et al.:  JCAP {\bf 0810}  045  : arXiv:0807.2575 [hep-th](2008)\\
Bamba, K., et al.:  Eur.\ Phys.\ J.\ C {\bf 67} 295: arXiv:0911.4390 [hep-th] (2010) \\
Bamba, K., et al.:Europhys.\ Lett.\  {\bf 89} 50003 : arXiv:0909.4397 [hep-th](2010) \\
Bamba, K., et al.:
  JCAP010,08 : arXiv:1309.3748 [hep-th](2014)\\
Bamba, K., et al.:
  Phys.\ Lett.\ B {\bf 732}, 349: arXiv:1403.3242 [hep-th](2014)
Barrow, J., Rliddle, A., Pahud, C.: Phys. Rev. D {\bf 74}, 127305 (2006)\\
Bennett, C.L., et al.: Astrophys. J. 583, 1 (2003)\\
Bhadra, J., Debnath, U.: Eur. Phys. J. C \textbf{72}, 1912 (2012)\\
Cai,Y., et al.:Phys.\ Rept.\  {\bf 493}, 1: arXiv:0909.2776 [hep-th](2010)\\
Cai,Y., et al.:Class.\ Quant.\ Grav.\  {\bf 28}, 215011 : arXiv:1104.4349 [astro-ph.CO] (2011)\\
Capozziello,S.,  et al.:
  Phys.\ Rev.\ D {\bf 87}, no. 8, 084037  : arXiv:1302.0093 [gr-qc](2013)\\
Capozziello,S.,et al.:
  Phys.\ Lett.\ B {\bf 671}  193  : arXiv:0809.1535 [hep-th](2009) \\
Cognola, G.,et al.:
  Eur.\ Phys.\ J.\ C {\bf 64}  483: arXiv:0905.0543 [gr-qc] (2009)\\
Cognola, G.,et al.:
  Phys.\ Rev.\ D {\bf 73} 084007  :  hep-th/0601008  (2006)\\
Cognola, G.,et al.:
  Phys.\ Rev.\ D {\bf 75} 086002  : hep-th/0611198  (2007)\\
Cognola, G.,et al.:
Phys.Rev. D77  ,046009: arXiv:0712.4017 [hep-th] (2008)\\
Clifton, T., et al.: Phys. Rep. 513, 1(2012)\\
Cvetic, M.,et al.:
  Nucl.\ Phys.\ B {\bf 628} 295: hep-th/0112045  (2002)\\
Elizalde ,E., et al.:Phys.Rev. D83 , 086006: arXiv:1012.2280 [hep-th] (2011)\\
Farooq, M. U., et al.:Can.\ J.\ Phys.\  {\bf 91} (2013) 703: arXiv:1306.1637 [astro-ph.CO] (2013)\\
Ferraro, R., Fiorini, F.: Phys. Rev. D 75, 084031 (2007)\\
Houndjo, M. J ., et al.:  Int.\ J.\ Mod.\ Phys.\ D {\bf 21}, 1250093 : arXiv:1206.3938 [physics.gen-ph] (2012)\\
Houndjo, M. J ., et al.:  arXiv:1304.1147 [physics.gen-ph] (2013)\\
Jamil, M.: Eur. Phys. J. C \textbf{62}, 325 (2009)\\
Jamil, M., Qadir, A.: Gen. Relativ. Gravit. \textbf{43}, 1069 (2011)\\
Jamil, M., Rashid, M.A., Qadir, A.: Eur. Phys. J. C \textbf{58}, 325 (2008)\\
Jamil, M., et al.:  arXiv:1309.3269 [gr-qc](2013)\\
Jamil, M., et al.:  Eur.\ Phys.\ J.\ C {\bf 72}  2267: arXiv:1212.6017 [gr-qc] (2012)\\
Jamil, M., et al.:  J.\ Phys.\ Soc.\ Jap.\  {\bf 81}  114004 : arXiv:1211.0018 [physics.gen-ph] (2012)\\
Jamil, M., et al.:Gen.\ Rel.\ Grav.\  {\bf 45}, 263: arXiv:1211.3740 [physics.gen-ph] (2013)\\
Jamil, M., et al.:  Eur.\ Phys.\ J.\ C {\bf 72}, 2137: arXiv:1210.0001 [physics.gen-ph] (2012)\\
Jamil, M., et al.:  Eur.\ Phys.\ J.\ C {\bf 72}, 2122 : arXiv:1209.1298 [gr-qc] (2012)\\
Jamil, M., et al.:  Chin.\ Phys.\ Lett.\  {\bf 29}, 109801: arXiv:1209.2916 [physics.gen-ph] (2012)\\
Jamil, M., et al.:  Eur.\ Phys.\ J.\ C {\bf 72}, 2075 : arXiv:1208.0025 (2012)\\
Jamil, M., et al.:Central Eur.\ J.\ Phys.\  {\bf 10}, 1065 : arXiv:1207.2735 [gr-qc] (2012)\\
Jamil, M., et al.:  Eur.\ Phys.\ J.\ C {\bf 72}, 1959 : arXiv:1202.4926 [physics.gen-ph](2012)\\
Jamil, M., et al.:Eur.\ Phys.\ J.\ C {\bf 71}, 1711: arXiv:1107.1558 [physics.gen-ph](2011)\\
Jamil, M., et al.:Eur.\ Phys.\ J.\ C {\bf 72}, 1999  : arXiv:1107.5807 [physics.gen-ph] (2012)\\
Kofinas,G., Saridakis, E. N.: arXiv:1404.2249 [gr-qc] (2014)\\
Kofinas, G., et al.: arXiv:1404.7100 [gr-qc] (2014)\\
Li, M., et al.: Commun. Theor. Phys. 56, 525 (2011)\\
Li, M.: Phys. Lett. B, 603, 1 (2004)\\
Lidsey,J. E.,
  JHEP {\bf 0206}  026   : hep-th/0202198  (2002)\\
Lidsey,J. E.,
  Phys.\ Lett.\ B {\bf 544}, 337  : hep-th/0207009 (2002)\\
Makarenko, A. N.,et al.:
 : arXiv:1403.7409 [hep-th]\\
Martin-Moruno, P.: Phys. Lett. B \textbf{659}, 40 (2008)\\
Momeni, D., Setare, M. R.Mod.\ Phys.\ Lett.\ A {\bf 26}, 2889 : arXiv:1106.0431 [physics.gen-ph](2011)\\
Momeni, D.,et al.:Europhys.\ Lett.\  {\bf 97}, 61001 : arXiv:1204.1246 [hep-th](2012)\\
Myrzakulov, R., Sebastiani, L.: Astrophys.\ Space Sci.\  {\bf 352}, 281 : arXiv:1403.0681 [gr-qc] (2014)\\
Nojiri, S., Odintsov, S.D.: Phys. Rev. D {\bf 74}, 086005 (2006)\\
Nojiri, S., Odintsov, S.D.: eConf C {\bf 0602061} (2006) 06
   [Int.\ J.\ Geom.\ Meth.\ Mod.\ Phys.\  {\bf 4}  115] : hep-th/0601213 (2007)\\
Nojiri, S., Odintsov, S.D.:  Gen. Rel. Grav. 38, 1285  [hep-th/0506212](2006).\\
Nojiri, S., Odintsov, S.D.:Int.\ J.\ Geom.\ Meth.\ Mod.\ Phys.\  {\bf 11}, 1460006: arXiv:1306.4426 [gr-qc] (2014)\\
Nojiri, S., Odintsov, S.D.:
  Phys.\ Rev.\ D {\bf 68}, 123512   : hep-th/0307288 (2003)\\
Nojiri, S., et al.:
  Gen.\ Rel.\ Grav.\  {\bf 42}  1997 : arXiv:0912.2488 [hep-th](2010)\\
Nojiri, S., et al.:
  Int.\ J.\ Mod.\ Phys.\ A {\bf 18}, 3395 : hep-th/0212047(2003)\\
Nojiri, S., Odintsov, S.D.:
 Phys.Lett. B657  ,238: arXiv:0707.1941 [hep-th] (2007)\\
Nojiri, S., et al.:
  Phys.\ Lett.\ B {\bf 651}  224  : arXiv:0704.2520 [hep-th] (2007)\\
Nojiri, S., Odintsov, S.D.:
  J.\ Phys.\ Conf.\ Ser.\  {\bf 66} 012005  : hep-th/0611071 (2007) \\
Nojiri, S., et al.:
  Phys.\ Rev.\ D {\bf 74}  046004  : hep-th/0605039 (2006)\\
Nojiri, S., et al.:
  J.\ Phys.\ A {\bf 39}  6627 : hep-th/0510183 (2006)\\
Nojiri, S., Odintsov, S.D.:
  Phys.\ Lett.\ B {\bf 631}, 1   : hep-th/0508049 (2005)\\
Nojiri, S., et al.:
  Phys.\ Rev.\ D {\bf 71} 123509  : hep-th/0504052  (2005)\\
Nojiri, S., Odintsov, S.D.:
  Gen.\ Rel.\ Grav.\  {\bf 37} 1419  : hep-th/0409244  (2005) \\
Nojiri, S., et al.:
  Int.\ J.\ Mod.\ Phys.\ A {\bf 17} 4809
  : hep-th/0205187 (2002) \\
Nojiri, S., Odintsov, S.D.: Phys. Lett. B 631, 1 : hep-th/0508
049 (2005) \\
Nojiri, S., Odintsov, S.D.:,Phys. Rev. D 74, 086005  : hep-th/0
608008 (2006)\\
Nojiri, S., Odintsov, S.D.: J. Phys. Conf. Ser. 66, 012005
: hep-th/0611071 (2007)\\
Nojiri, S., et al.: Phys. Le
tt. B
681
, 74  : arXiv:0908.1269
[hep-th] (2009)\\
Nojiri, S., Odintsov, S.D.:
  Phys.\ Rept.\  {\bf 505}  59
  : arXiv:1011.0544 [gr-qc] (2011)\\
Nojiri, S., Odintsov, S.D.:
 Phys.Rev. D78 ,046006
: arXiv:0804.3519 [hep-th]  (2008)\\
 Odintsov, S.D. et al.:
 :1406.1205 [hep-th] (2014)\\
Padmanabhan, T.: Curr. Sci. 88, 1057 (2005)\\
Perlmutter, S., et al.: Astrophys. J. 517, 565 (1999)\\
Rodrigues, M. E., et al.: JCAP {\bf 1311} 024: arXiv:1306.2280 [gr-qc] (2013)\\
Rodrigues, M. E., et al.: Int.\ J.\ Mod.\ Phys.\ D {\bf 22}  8,  1350043 : arXiv:1302.4372 [physics.gen-ph] (2013)\\
Sahni, V., Starobinsky, A.: Int. J. Mod. Phys. D 15, 2105 (2006)\\
Setare, M. R.,  Momeni, D.Int.\ J.\ Theor.\ Phys.\  {\bf 50}, 106: arXiv:1001.3767 [physics.gen-ph](2011)\\
Sharif, M., Abbas, G.: Chin. Phys. Lett. \textbf{28}, 090402 (2011)\\
Sharif, M., Jawad, A.: Int. J. Mod. Phys. D \textbf{22}, 1350014 (2013)\\
Sharif, M., Jawad, A.: Eur. Phys. J. C \textbf{73}, 2382 (2013a)\\
Sharif, M., Jawad, A.: Eur. Phys. J. C \textbf{73}, 2600 (2013b)\\
Sharif, M., Jawad, A.: Eur. Phys. J. Plus \textbf{129}, 15 (2014)\\
Sharif, M., Rani, S.: Astrophy. Space Sci. \textbf{345}, 217
(2013)\\
Sharif, M., Zubair, M.: Astrophysics and Space Science, 1 (2014)\\
Spergel, D.N., et al.: Astrophys. J. Suppl. Ser. 148, 175 (2003)\\
Tegmark, M., et al.: Phys. Rev. D 69, 103501 (2004)\\
Tsujikawa, S.: Lect. Notes Phys. 800, 99 (2010)\\
Wei, H.: Classical and Quantum Gravity, 29, 175008 (2012).
\end{document}